\documentclass[twocolumn,10pt]{IEEEtran}
\usepackage{amssymb}
\usepackage{theorem,amsbsy,pifont,verbatim,amsfonts,amssymb,color}
\usepackage{xspace,epsfig,graphicx,subfigure,latexsym,fancyhdr,cite}
\usepackage{multirow}
\usepackage{epstopdf}
\usepackage{makecell}
\usepackage{bm}
\usepackage{booktabs}
\usepackage{indentfirst}
\usepackage{amsmath}
\usepackage{diagbox}
\usepackage{enumerate}
\usepackage{float}
\usepackage{url}
\usepackage{mdwmath}
\usepackage{mdwtab}
\usepackage{amsmath}
\usepackage{xcolor}
\usepackage{caption}
\usepackage{graphfig}
\usepackage{algorithm}
\usepackage{algpseudocode}
\usepackage[colorlinks,citecolor=blue,linkcolor=blue,urlcolor=blue]{hyperref}

\newtheorem{remark}{Remark}
 
\allowdisplaybreaks

\ifCLASSINFOpdf

\else

\fi

\usepackage{stfloats}

\begin{document}
	
	\title{Scalable Fluid Antenna Systems: \\A New Paradigm for Array  Signal Processing}
	
	\author{Tuo Wu, Ye Tian, Jie Tang, Kangda Zhi,  Maged Elkashlan,  \\  Kin-Fai Tong, \emph{Fellow, IEEE},    Naofal Al-Dhahir, \emph{Fellow, IEEE},   \\   Chan-Byoung Chae, \emph{Fellow, IEEE},  Matthew  C. Valenti, \emph{Fellow}, \emph{IEEE}, \\ ~George K. Karagiannidis,~\IEEEmembership{Fellow,~IEEE},
		and		 Kwai-Man Luk, \emph{Life Fellow, IEEE}
		
		\thanks{(\textit{Corresponding author: Ye Tian.})}
		\thanks{This research work of T. Wu was funded by Hong Kong Research Grants Council under the Area of Excellence Scheme under Grant AoE/E-101/23-N.
			T. Wu and K.-M. Luk are with the State Key Laboratory of Terahertz and
			Millimeter Waves, Department of Electronic Engineering, City University of
			Hong Kong, Hong Kong. (E-mail: $\rm \{tuowu2,eekmluk\}@cityu.edu.hk$).
			Y. Tian is with the Faculty of Electrical Engineering and Computer Science, Ningbo University, Ningbo 315211, China (E-mail: $\rm tianye1@nbu.edu.cn$).
			J. Tang is with the School of Electronic and Information Engineering, South China University of Technology, Guangzhou 510640, China (E-mail: $\rm eejtang@scut.edu.cn$).
			K. Zhi is with Communications and Information Theory Group (CommIT), Technische Universitat Berlin, 10587 Berlin, Germany (E-mail: $\rm k.zhi@tu$-$\rm berlin.de$).
			M. Elkashlan is with the School of Electronic Engineering and Computer Science at Queen Mary University of London, London E1 4NS, U.K. (E-mail: $\rm maged.elkashlan@qmul.ac.uk$). 
			Naofal Al-Dhahir is with the Department of Electrical and Computer Engineering, The University of Texas at Dallas, Richardson, TX 75080 USA (E-mail: $\rm aldhahir@utdallas.edu$).   
			K. F. Tong is with the School of Science and Technology, Hong Kong Metropolitan University, Hong Kong SAR, China. (E-mail: $\rm  ktong@hkmu.edu.hk$).
			C.-B. Chae is with the School of Integrated Technology, Yonsei University, Seoul 03722 Korea. (E-mail: $\rm cbchae@yonsei.ac.kr$). 
			M. C. Valenti is with the Lane Department of Computer Science and Electrical Engineering, West Virginia University, Morgantown, USA (E-mail: $\rm valenti@ieee.org$).
			G. K. Karagiannidis is with the Department of Electrical and Computer Engineering, Aristotle University of Thessaloniki, 54124 Thessaloniki, Greece (E-mail: $\rm geokarag@auth.gr$).
	}}
	
	\maketitle
	
	\begin{abstract}
		Most existing antenna array-based source localization methods rely on fixed-position arrays (FPAs) and strict assumptions about source field conditions (near-field or far-field), which limits their effectiveness in complex, dynamic real-world scenarios where high-precision localization is required. In contrast, this paper introduces a novel scalable fluid antenna system (SFAS) that can dynamically adjust its aperture configuration to optimize performance for different localization tasks. Within this framework, we develop a two-stage source localization strategy based on the exact spatial geometry (ESG) model: the first stage uses a compact aperture configuration for initial direction-of-arrival (DOA) estimation, while the second stage employs an expanded aperture for enhanced DOA and range estimation. The proposed approach eliminates the traditional need for signal separation or isolation to classify source types and enables a single SFAS array to achieve high localization accuracy without field-specific assumptions, model simplifications, or approximations, representing a new paradigm in array-based source localization. Extensive simulations demonstrate the superiority of the proposed method in terms of localization accuracy, computational efficiency, and robustness to different source types.
		
	\end{abstract}
	
	
	\section{Introduction}
	
	\IEEEPARstart{D}{irection}-of-arrival (DOA) estimation plays a crucial role in a multitude of military and civilian applications. In wireless communications, accurate DOA estimation is essential for acquiring channel state information (CSI) and performing effective downlink beamforming \cite{ref1,rr1,ref2}, while also enhancing target detection and tracking capabilities in radar and sonar systems \cite{ref3}. To date, a plethora of excellent DOA estimation methods have been proposed, including subspace-based approaches like multiple signal classification (MUSIC) \cite{ref4} and rank-reduction (RARE) \cite{ref5}, sparse signal reconstruction methods such as sparse Bayesian learning (SBL) \cite{ref6}, and deep learning-based solutions \cite{ref8,rr7,ref9}.
	
	However, these established methods are fundamentally built upon fixed-position antennas (FPAs) with element spacing typically no larger than half the carrier wavelength. Such a rigid architecture inherently suffers from two major limitations. First, the relatively wide element spacing is to avoid mutual coupling effects, which severely degrades estimation performance \cite{ref10}. Second, the static steering vector of an FPA corresponds to a fixed number of degrees-of-freedom (DoFs), restricting the ability to achieve super-resolution and resolve a large number of sources. While massive multiple-input multiple-output (MIMO) arrays can partially address these issues, their high hardware costs and power consumption are often prohibitive.
	
	As an alternative solution, fluid antenna systems (FASs),  have emerged as a promising approach in recent years, offering the potential to overcome the limitations of FPA systems \cite{ref11,refnew11,refnew12,ref12,ref13,ref14,LaiX242,YaoJ241,XLai23}. In an FAS, the position of each radiating element can be dynamically reconfigured across a spatial aperture, enabling flexible and adaptive control over the antenna's spatial behavior. This reconfigurability can be achieved through various means, such as electronically switchable pixel arrays, metasurfaces, or other tunable structures, without necessarily requiring any physical displacement or fluidic materials. By allowing the effective radiation point to change in response to the environment or communication needs, FA systems unlock additional spatial DoFs, offering new opportunities for enhancing performance in next-generation wireless systems. Motivated by these advantages, extensive research has explored FA-enabled schemes, including fluid antenna multiple access (FAMA) \cite{add13,add14,add15}, channel estimation \cite{add16,add17,add18,	aadd18}, beamforming design \cite{add19,add20}, and integrated sensing and communications (ISAC) \cite{add21,add22,	JYao2024,YaoJ252,Zheng2025,Tang2025,Ghadi2024,Yangsx2025,Shojaeifard,Huang21,Rodrigo14,Waqar23}.

	\begin{figure*}[t]  \centering
		\includegraphics[width=6.5in]{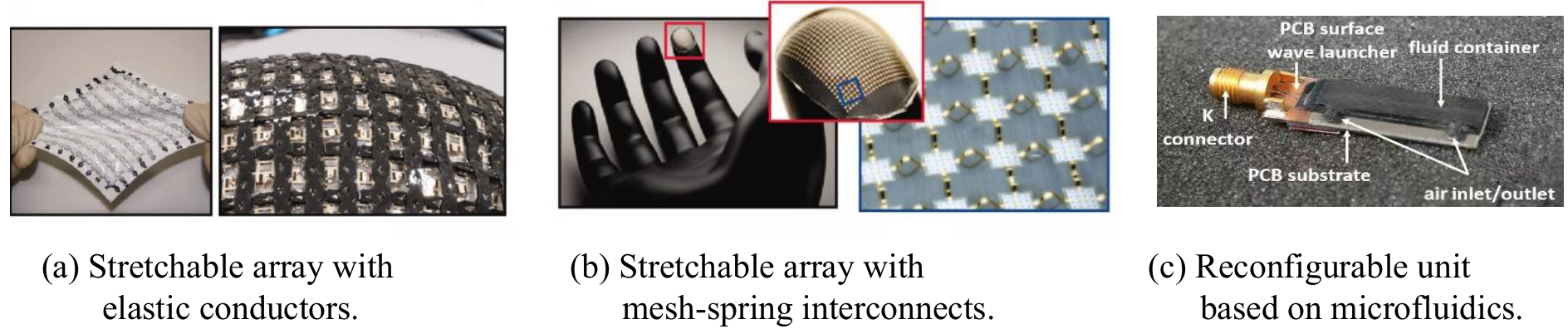}
		\caption {Conceptual prototypes for the physical realization of S-FAS, inspired by state-of-the-art research in stretchable electronics and microfluidics. (a) A stretchable array realized with intrinsically elastic conductors connecting rigid components \cite{rogers2010materials}. (b) A mesh-spring design where stretchability is achieved through the deformation of spring-like interconnects between rigid nodes \cite{rogers2010materials}. (c) A single reconfigurable antenna unit where characteristics are altered by controlling the position of a conductive liquid within a microfluidic channel \cite{Huang21}, forming the basis for a scalable array via selective channel activation.\textit{These prototypes are conceptual illustrations; this paper does not report hardware measurements but abstracts the implementation-driven constraints into modeling assumptions and robustness analyses.}}\label{system}			\vspace{-3mm}
	\end{figure*}

	Despite FAS's demonstrated versatility, its potential for enhancing direction-finding capabilities remains largely unexplored. The unique characteristics of FAS offer several key advantages for DOA estimation. First, the dynamic movement of the antenna can construct a larger virtual array, significantly increasing spatial DoFs. This not only enhances estimation accuracy but also enables underdetermined DOA estimation, where the number of detectable sources exceeds the number of physical antennas. Second, the flexible positioning allows for adaptive array configurations optimized for different scenarios, providing superior spatial resolution.

	Nevertheless, even with the benefits of FAS, a fundamental challenge persists, inherited from traditional array signal processing. The core motivation for this work stems from the recognition that conventional array signal processing faces three fundamental limitations when applied to realistic localization scenarios. First, \emph{rigid geometric constraints} force a compromise between enhancing spatial resolution (favoring large spacing) and avoiding grating lobe ambiguities  (favoring small spacing), preventing optimal adaptation to different estimation scenarios. Here, ``rigid geometric constraints" refers to the fact that traditional arrays are locked into a fixed inter-element spacing once manufactured, forcing engineers to make permanent compromises between conflicting design requirements. Second, \emph{field classification dependence} requires a priori knowledge of whether sources are in near-field or far-field regions. Most localization methods are built on rigid field-specific assumptions: for sources in the \emph{far-field}, defined by the Fraunhofer distance $r \gg \frac{2D^2}{\lambda}$ (where $D$ is the array aperture and $\lambda$ is the wavelength), methods rely on the computationally simple planar wavefront approximation; for sources in the \emph{near-field}, typically within the Fresnel region $0.62(D^3/\lambda)^{1/2} < r < 2D^2/\lambda$, they employ a more complex spherical wavefront model, often simplified using a second-order Taylor approximation. This leads to suboptimal performance when this classification is incorrect or when mixed-field scenarios are encountered. 
	
	These limitations become particularly acute in emerging 6G applications where multiple sources may simultaneously exist across near-field, Fresnel, and far-field regions, creating complex mixed-field environments that traditional methods cannot handle effectively. The strict reliance on predefined field regions and the model mismatch introduced by approximation-based approaches result in systematic estimation errors that degrade localization accuracy   when high precision is most critical.

	To address these challenges, we introduce a new paradigm: \textbf{scalable FAS (S-FAS)}, which represents a fundamental paradigm shift from field-specific optimization to universal, adaptive localization. The core innovation of S-FAS lies in its ability to dynamically \textbf{scale} its physical aperture through a software-controlled scaling mechanism. This scalability is not merely a theoretical construct but is grounded in various advanced technologies, as illustrated by the conceptual prototypes in Fig.~\ref{system}. As shown in Fig.~\ref{system}(a) and Fig.~\ref{system}(b), one approach, inspired by stretchable electronics, is to use structural designs like intrinsically elastic conductors or mesh-spring interconnects to physically stretch and compress the array \cite{rogers2010materials}. Another approach, depicted in Fig.~\ref{system}(c), employs microfluidics, where the positions of active elements can be reconfigured by controlling liquid within a channel network \cite{Huang21}.
	
	This unique, physically-grounded scaling capability enables S-FAS to escape the fundamental design trade-offs that plague traditional systems, creating a powerful, adaptive localization framework. The system operates by intelligently switching between two complementary configurations: when compressed to sub-wavelength spacing, it creates a compact array that completely eliminates grating lobe ambiguities, though mutual coupling becomes stronger, this compact geometry is ideal for obtaining reliable initial direction estimates without angular confusion. Conversely, when expanded to super-wavelength spacing, the system generates a large-aperture array with exceptional spatial resolution and negligible mutual coupling, perfect for precise refinement of both angle and range parameters. Through this adaptive approach, S-FAS transcends the traditional compromise between robustness and precision, instead achieving both objectives sequentially within a unified framework optimized for different localization objectives.
	
	However, this flexibility introduces new challenges that demand innovative solutions. First, the elimination of field classification dependence requires developing a \emph{unified channel model} that can accurately characterize signal propagation across all field regimes without approximations—from near-field spherical wavefronts to far-field planar waves and everything in between. Second, the compressed configuration introduces \emph{severe mutual coupling challenges} due to the sub-wavelength spacing, necessitating sophisticated coupling compensation techniques that can extract useful signal information despite strong electromagnetic interference between closely-spaced elements. Third, when sources exist in \emph{intermediate field regions} (such as the Fresnel zone), traditional far-field and near-field models both fail, requiring new approaches to handle these transitional scenarios effectively. Finally, the expanded configuration demands \emph{high-precision joint estimation algorithms} that can simultaneously refine both angle and range parameters while exploiting the enhanced spatial resolution of the large aperture, presenting significant computational and algorithmic design challenges.
	
	We systematically address each of these challenges through a novel two-stage estimation strategy built upon an exact spatial geometry (ESG) model. Our approach strategically decomposes the complex mixed-field localization problem into manageable stages, with each stage specifically designed to overcome the corresponding technical barriers: a compressed configuration that tackles mutual coupling and grating lobe issues for robust initial DOA estimation, and an extended configuration that exploits enhanced spatial resolution for precise joint angle-range refinement in all field regimes. The main contributions of this work are summarized as follows:
	
	\begin{itemize}
		\item \textbf{\textit{Novel S-FAS Architecture:}} We develop the first scalable fluid antenna system specifically designed for mixed-field source localization, featuring software-controlled dynamic aperture reconfiguration capabilities. Our system fundamentally eliminates the rigid geometric constraints and field classification dependence that plague traditional approaches.
		
		\item \textbf{\textit{Unified Exact Spatial Geometry Model:}} We establish the first comprehensive mathematical framework based on exact spatial geometry that provides approximation-free unified handling of near-field, Fresnel, and far-field sources within a single model. This breakthrough eliminates the need for field classification while naturally incorporating range information as a first-class parameter, gracefully transitioning from joint angle-range estimation to pure DOA estimation based on source characteristics.
		
		\item \textbf{\textit{Adaptive Mutual Coupling Mitigation:}} We develop sophisticated mutual coupling compensation techniques specifically tailored for dynamic array configurations, including  rank-reduction methods for super-wavelength extended configurations, ensuring robust signal extraction despite severe electromagnetic interference.
		
		\item \textbf{\textit{Two-Stage Estimation Framework:}} We propose an innovative decomposition strategy that strategically exploits the complementary advantages of different array configurations through intelligent problem partitioning. The compressed stage provides grating-lobe-free initial estimates despite strong coupling, while the extended stage achieves high-precision joint refinement with enhanced spatial resolution and negligible coupling.
		
		\item \textbf{\textit{Comprehensive Performance Validation:}} Through extensive simulations across diverse scenarios—including challenging mixed-field environments with sources spanning near-field (30$\lambda$), Fresnel region (300$\lambda$), and far-field (5000$\lambda$) distances—we demonstrate that our S-FAS framework achieves centimeter-level positioning accuracy while maintaining computational efficiency, outperforming state-of-the-art methods by significant margins across all field regimes.
	\end{itemize}
	
	
	The remainder of this paper is organized as follows. Section~\ref{sec:system_overview} provides a comprehensive overview of the S-FAS architecture and signal models. Section~\ref{sec:algorithms} presents the proposed two-stage estimation algorithms with mutual coupling mitigation techniques. Section~\ref{sec:simulations} provides detailed simulation results demonstrating the framework's effectiveness. Finally, Section~\ref{sec:conclusion} concludes the paper with discussions on future research directions.
				\vspace{-3mm}
	\section{System Overview}\label{sec:system_overview}

	\subsection{Conventional Array Signal Models: Limitations and Motivations}
	To motivate the development of our proposed scalable fluid antenna system (S-FAS), we first examine the fundamental limitations of conventional array signal processing approaches. Traditional methods in mixed-field scenarios rely on either far-field or near-field approximations, each imposing significant constraints on estimation performance and applicability. Moreover, practical implementation factors such as mutual coupling effects further complicate the signal model, especially in compact array configurations.
	
	\subsubsection{Traditional Far-Field Model}
	Consider a conventional uniform linear array (ULA) with $M$ antenna elements and inter-element spacing $d = \lambda/2$, where $\lambda$ denotes the carrier wavelength. For a source located at angle $\theta$ and distance $r$ from the array center, the traditional far-field model assumes $r \gg \frac{2D^2}{\lambda}$ (where $D = (M-1)d$ is the array aperture), leading to the planar wavefront approximation.
	
	The array steering vector under far-field assumption is formulated as
	\begin{equation}\label{eq:far_field_steering}
		\mathbf{a}_{\text{FF}}(\theta) = [1, e^{-j\frac{2\pi}{\lambda}d\sin\theta}, e^{-j\frac{2\pi}{\lambda}2d\sin\theta}, \ldots, e^{-j\frac{2\pi}{\lambda}(M-1)d\sin\theta}]^T.
	\end{equation}
				\vspace{-3mm}
	\begin{remark}
		This {far-field model} introduces systematic errors when sources are not in the {traditional defined} far-field region, particularly for large arrays where the planar wavefront assumption becomes invalid.
	\end{remark}
	
	\subsubsection{Traditional Near-Field Model}
	For near-field scenarios where {$0.62(D^3/\lambda)^{1/2}<r < {2D^2/\lambda}$}, the exact distance from source to each antenna element must be considered. The distance from a source at $(\theta, r)$ to the $m$-th antenna element is expressed as
	\begin{equation}\label{eq:near_field_distance}
		r_{m} = \sqrt{r^2 + [(m-1)d]^2 - 2r(m-1)d\sin\theta},
	\end{equation}
	{by utilizing the second-order Taylor expansion approximation, the near-field steering vector can be approximately written as}
	\begin{align}\label{eq:near_field_steering}
		\mathbf{a}_{\text{NF}}(\theta, r)=& [1, e^{-j\frac{2\pi}{\lambda}d\sin\theta+j(\frac{\pi d^2}{\lambda r}\cos^2\theta)}, \nonumber\\
		& e^{-j\frac{2\pi}{\lambda}(M-1)d\sin\theta+j((M-1)^2\frac{\pi d^2}{\lambda r}\cos^2\theta)}]^T.
	\end{align}
	
	\begin{remark}
		{This near-field model still introduces systematic errors due to the second-order approximation. Moreover, when sources approach the Fresnel boundary or are beyond the Fresnel region, such a signal model is invalid and effective localization performance cannot be guaranteed. Besides, it is noted that traditional near-field processing normally requires joint estimation of both angle and range, and leading to high computational complexity.}
	\end{remark}
	
	\subsubsection{Mutual Coupling Effects in Practical Arrays}
	A critical limitation often overlooked in conventional array signal processing is the presence of mutual coupling between antenna elements. When antenna elements are positioned close to each other, the electromagnetic field radiated by one element induces currents in neighboring elements, creating parasitic coupling effects that distort the ideal array response.
	
	For conventional arrays with inter-element spacing $d \leq \lambda/2$, mutual coupling becomes significant and cannot be ignored. The coupling-affected steering vector is expressed as
	\begin{equation}\label{eq:coupling_affected_steering}
		\mathbf{a}_{\text{c}}(\theta) = \mathbf{C} \mathbf{a}_{\text{i}}(\theta),
	\end{equation}
	where $\mathbf{C} \in \mathbb{C}^{M \times M}$ is the mutual coupling matrix with Toeplitz structure, and $\mathbf{a}_{\text{i}}(\theta)$ represents the ideal steering vector (far-field or near-field).
	
	The mutual coupling matrix typically follows the structure
	\begin{equation}\label{eq:coupling_matrix_conventional}
		\mathbf{C} = \begin{bmatrix}
			1 & c_1 & c_2 & \cdots & c_{M-1} \\
			c_1^* & 1 & c_1 & \cdots & c_{M-2} \\
			c_2^* & c_1^* & 1 & \cdots & c_{M-3} \\
			\vdots & \vdots & \vdots & \ddots & \vdots \\
			c_{M-1}^* & c_{M-2}^* & c_{M-3}^* & \cdots & 1
		\end{bmatrix},
	\end{equation}
	where $c_m$ represents the coupling coefficient between elements separated by $m$ positions. These coefficients are determined by the physical spacing and electromagnetic properties of the array.
	
	\begin{remark}
		Mutual coupling effects become more pronounced as inter-element spacing decreases, leading to significant performance degradation in compact array configurations. Traditional compensation methods require offline calibration and assume fixed array geometry, making them unsuitable for adaptive systems.
	\end{remark}
	
	\subsubsection{Motivation for S-FAS Approach}
	The analysis of conventional array signal models reveals fundamental limitations that motivate our S-FAS solution. Traditional arrays suffer from rigid geometric constraints, where a fixed inter-element spacing forces a permanent trade-off between avoiding grating lobes (which favors smaller spacing of $d \leq \lambda/2$) and avoiding mutual coupling (which favors larger spacing), preventing optimal adaptation to diverse scenarios. Furthermore, they exhibit a strong field classification dependence, requiring a priori knowledge of whether sources are in near-field or far-field regions, leading to significant performance degradation from model mismatch in complex mixed-field environments. Finally, existing static coupling compensation methods are designed for fixed geometries and cannot adapt to the dynamic spacing changes inherent in a reconfigurable system. \textit{These challenges highlight the urgent need for a more flexible and adaptive localization framework.}
	
	Despite these achievements, the current S-FAS framework has several limitations that provide clear directions for future research. \textit{First}, our approach assumes that source positions remain stationary during the transition from the compressed to extended configuration. In practical scenarios, mobile sources may change location between the two stages, potentially degrading estimation accuracy. Future work should investigate robust tracking algorithms that can handle source mobility during configuration transitions, possibly through predictive filtering or iterative refinement strategies. \textit{Second}, the current framework is limited to 2D positioning, estimating only range and a single azimuth angle for sources in a coplanar geometry. Extension to full 3D positioning would require estimation of both azimuth and elevation angles, necessitating planar or volumetric S-FAS arrays and corresponding signal models. \textit{Third}, while our mutual coupling mitigation techniques are effective, they add computational overhead that may limit real-time applications. Future research could explore machine learning-based coupling compensation or hardware-level coupling reduction techniques. Additionally, investigating the scalability of S-FAS to larger array sizes and multiple simultaneous reconfigurations represents an important avenue for enhancing system capacity and flexibility in future sixth-generation wireless communication systems.
	
	
	
	
	
	
	\begin{figure}[t]  \centering
		\includegraphics[width=3.4in]{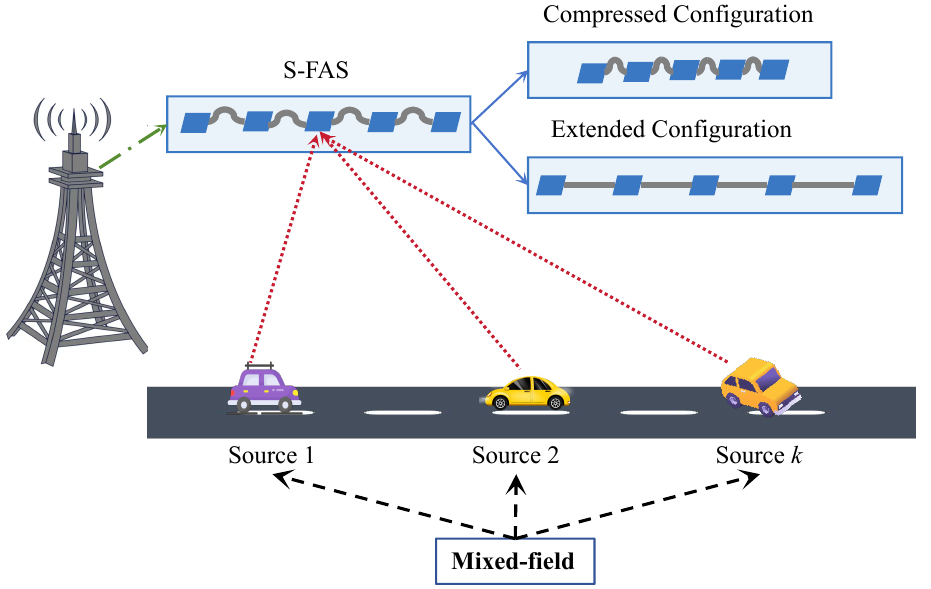}
		\caption {System model of S-FAS for mixed-field sources  localization system.}\label{system_model}			\vspace{-3mm}
	\end{figure}
				\vspace{-3mm}
	\subsection{Scalable Fluid Antenna System (S-FAS) Design}
	To address the fundamental limitations of conventional models discussed above, we propose an S-FAS, a reconfigurable antenna array designed for high-precision localization in challenging mixed-field environments. As illustrated in the system model in Fig.~\ref{system_model}, an S-FAS can be deployed at a base station to locate multiple sources, such as vehicles, which may simultaneously occupy near-field, Fresnel, and far-field regions. The fundamental innovation of S-FAS lies in its unique capability to dynamically switch between a \textit{compressed configuration}  and an \textit{extended configuration}.  Throughout the estimation process, source locations are assumed to remain stationary, enabling coherent processing across both configurations.
	
	
	\subsubsection{S-FAS Configuration Parameters}
	Consider an S-FAS consisting of $M$ antenna elements whose positions can be dynamically controlled. The key innovation lies in the ability to adjust the inter-element spacing according to a scaling factor $\alpha$, which is given by
	\begin{equation}\label{eq:scaling_spacing}
		d(\alpha) = \alpha \cdot d_0,
	\end{equation}
	where $d_0 = \frac{\lambda}{2}$ is the baseline inter-element spacing, $\alpha \in [\alpha_{\min}, \alpha_{\max}]$ is the scaling factor controlling aperture configuration. For this work, we consider a dual-configuration approach with $\alpha \in \{\alpha_c, \alpha_e\}$, where $\alpha_c < 1$ represents a compressed configuration and $\alpha_e > 1$ represents an extended configuration.
	
	Using the scaling relationship in \eqref{eq:scaling_spacing}, the position of the $m$-th antenna element is formulated as
	\begin{equation}\label{eq:element_position}
		p_m(\alpha) = (m-1) \cdot d(\alpha) = (m-1) \cdot \alpha \cdot d_0, \quad m = 1, 2, \ldots, M
	\end{equation}
	consequently, the total array aperture is expressed as
	\begin{equation}\label{eq:array_aperture}
		D(\alpha) = (M-1) \cdot d(\alpha) = (M-1) \cdot \alpha \cdot d_0.
	\end{equation}
	The Rayleigh distance, which determines the near-field/far-field boundary, is configuration-dependent and derived from \eqref{eq:array_aperture} as
	\begin{equation}\label{eq:rayleigh_distance}
		R_F(\alpha) = \frac{2D^2(\alpha)}{\lambda} = \frac{2[(M-1)\alpha d_0]^2}{\lambda} = \alpha^2 \cdot R_{F0},
	\end{equation}
	where $R_{F0} = \frac{2[(M-1)d_0]^2}{\lambda}$ is the Rayleigh distance for the baseline configuration.
	
	\subsubsection{Two-Stage Estimator}
	To exploit the complementary advantages of different array geometries, the S-FAS employs a systematic two-stage estimator. The system collects $N$ snapshots per configuration and operates under additive white Gaussian noise (AWGN) with variance $\sigma^2_n$.
	
	\paragraph{Stage 1: Initial DOA Estimation with Compressed Configuration ($\alpha_c < 1$)}
	The first stage focuses on obtaining robust initial DOA estimates. This is achieved by operating the S-FAS in a compressed configuration, where the inter-element spacing $d_c = \alpha_c d_0 < \lambda/2$. The reduced array aperture, $D_c = (M-1)\alpha_c d_0$, \textit{allows for the use of the computationally efficient far-field approximation} \footnote{It is worth noting that even for sources located in the intermediate field (i.e., beyond the Fresnel region but not strictly satisfying the Fraunhofer condition), the initial DoA estimates obtained via this far-field approximation are typically close enough to the true values to ensure the convergence of the high-resolution search in Stage 2. The robustness of this approach is well-documented in the literature \cite{HXu24}.}.

	\paragraph{Stage 2: Range and DOA Refinement with Extended Configuration ($\alpha_e > 1$)}
	The second stage aims to refine the DOA estimates and perform precise range estimation. The S-FAS switches to an extended configuration, where the inter-element spacing is enlarged to $d_e = \alpha_e d_0$. This creates an expanded aperture, $D_e = (M-1)\alpha_e d_0$, which \textit{enhances spatial resolution.}
	
	
	This two-stage approach strategically balances the trade-offs between computational complexity, spatial resolution, and mutual coupling effects to achieve robust and accurate DoA estimation.
	
	\subsection{ Signal Model  for S-FAS}
	Having established the S-FAS architecture and design principles, we now develop a comprehensive signal model that captures the unique characteristics of mixed near-field and far-field source localization. The fundamental challenge addressed by our S-FAS approach is the elimination of the restrictive assumption that conventional methods require prior knowledge of source field classifications. Our proposed solution employs a unified  ESG  model as the theoretical foundation, while strategically implementing configuration-dependent processing for optimal performance.
	
	\subsubsection{ESG Model}
	The cornerstone of our S-FAS approach is the exact spatial geometry (ESG) model that provides a unified mathematical framework for characterizing propagation from arbitrary source locations. For the $k$-th source at $(\theta_k, r_k)$ and S-FAS configuration $\alpha$, the exact distance from the source to the $m$-th antenna element is given by
	\begin{equation}\label{eq:sfas_distance}
		r_{m,k}(\alpha) = \sqrt{r_k^2 + p_m^2(\alpha) - 2r_k p_m(\alpha) \sin\theta_k},
	\end{equation}
	accordingly, the ESG steering vector is formulated as
	\begin{align}\label{eq:sfas_steering}
		&\mathbf{a}_k^{\text{ESG}}(\theta_k, r_k, \alpha)\nonumber\\
		&= \left[\frac{r_k}{r_{1,k}(\alpha)}e^{j\phi_{1,k}(\alpha)},   \ldots, \frac{r_k}{r_{M,k}(\alpha)}e^{j\phi_{M,k}(\alpha)}\right]^T,
	\end{align}
	where the phase factor for the $m$-th element is expressed as
	\begin{equation}\label{eq:phase_factor}
		\phi_{m,k}(\alpha) = \frac{2\pi}{\lambda}[r_{m,k}(\alpha) - r_k].
	\end{equation}
	
	\begin{remark}
		The ESG model in \eqref{eq:sfas_steering} provides the exact propagation characteristics for all source locations regardless of their near-field or far-field classification. This unified model serves as the theoretical foundation for all subsequent processing stages.
	\end{remark}

	\subsubsection{Signal Model at Stage 1}
	In Stage 1, the S-FAS operates in a compressed configuration ($\alpha_c < 1$) to obtain initial DOA estimates. The received signal in this configuration is physically affected by mutual coupling acting on the true channel response. The mutual coupling matrix $\mathbf{C}_c(\alpha) \in \mathbb{C}^{M \times M}$, which follows a Toeplitz structure, modifies the ideal far-field response. The general form of the coupling coefficient for a given configuration $\alpha$ is expressed as
	\begin{equation}\label{eq:coupling_vector}
		c_m(\alpha) = c_0 \exp\left(-\beta \frac{m \cdot \alpha \cdot d_0}{\lambda}\right) e^{j\psi_m(\alpha)},
	\end{equation}
	where the phase component for configuration $\alpha$ is given by
	\begin{equation}\label{eq:coupling_parameters}
		\psi_m(\alpha) = \frac{2\pi m \cdot \alpha \cdot d_0}{\lambda} + \phi_0.
	\end{equation}
	Here, $c_0$ represents the reference coupling strength at baseline spacing, $\beta > 0$ is the coupling decay factor, and $\phi_0$ is a phase offset.
	
	The effective channel for the $k$-th source is given by the coupling matrix $\mathbf{C}_c(\alpha_c)$ acting on the ESG channel vector:
	\begin{equation}\label{eq:true_coupled_channel}
		\tilde{\mathbf{h}}_k(\alpha_c) = \sqrt{P_k} \mathbf{C}_c(\alpha_c) \mathbf{a}_k^{\text{ESG}}(\theta_k, r_k, \alpha_c).
	\end{equation}
	Due to the small array aperture in the compressed mode, the far-field assumption is valid for most sources. For computational efficiency, we can approximate the ESG steering vector with its far-field counterpart:
	\begin{equation}\label{eq:approx_coupled_channel}
		\tilde{\mathbf{h}}_k(\alpha_c) \approx \sqrt{P_k} \mathbf{C}_c(\alpha_c) \mathbf{a}_k^{\text{FF}}(\theta_k, \alpha_c),
	\end{equation}
	where the far-field steering vector is defined as
	\begin{equation}\label{eq:far_field_approx}
		\mathbf{a}_k^{\text{FF}}(\theta_k, \alpha) = [1, e^{-j\pi\alpha\sin\theta_k}, \ldots, e^{-j(M-1)\pi\alpha\sin\theta_k}]^T.
	\end{equation}
	
	Based on this approximation, the received signal vector at Stage 1 can be expressed as:
	\begin{equation}\label{eq:compact_signal_expanded}
		\mathbf{x}_c(t) \approx \mathbf{C}_c(\alpha_c) \mathbf{A}_{\text{FF}}(\theta, \alpha_c) \tilde{\mathbf{s}}(t) + \mathbf{n}_c(t).
	\end{equation}
	To separate the geometric factors from the signal powers, we define the far-field array manifold matrix $\mathbf{A}_{\text{FF}}(\theta, \alpha_c) \in \mathbb{C}^{M \times K}$ as
	\begin{equation}\label{eq:ff_manifold}
		\mathbf{A}_{\text{FF}}(\theta, \alpha_c) = [\mathbf{a}_1^{\text{FF}}(\theta_1, \alpha_c), \dots, \mathbf{a}_K^{\text{FF}}(\theta_K, \alpha_c)],
	\end{equation}
	and an effective source signal vector $\tilde{\mathbf{s}}(t) \in \mathbb{C}^{K \times 1}$ that incorporates the path gains:
	\begin{equation}\label{eq:effective_signal}
		\tilde{\mathbf{s}}(t) = [\sqrt{P_1}s_1(t), \dots, \sqrt{P_K}s_K(t)]^T.
	\end{equation}
	Substituting \eqref{eq:ff_manifold} and \eqref{eq:effective_signal} into \eqref{eq:compact_signal_expanded}, the signal model for Stage 1 simplifies to the compact form:
	\begin{equation}\label{eq:compact_signal}
		\mathbf{x}_c(t) \approx \mathbf{C}_c(\alpha_c) \mathbf{A}_{\text{FF}}(\theta, \alpha_c) \tilde{\mathbf{s}}(t) + \mathbf{n}_c(t).
	\end{equation}

	\subsubsection{Signal Model at Stage 2}
	
	In Stage 2, the S-FAS is reconfigured to an extended mode with a scaling factor $\alpha_e > 1$, resulting in an enlarged inter-element spacing $d_e = \alpha_e d_0$. A key advantage of this configuration is that the increased separation between elements renders mutual coupling effects negligible.
	
	Consequently, the signal can be modeled directly using the exact ESG framework from \eqref{eq:sfas_steering}. The channel matrix for the extended configuration, which includes the path gains, is
	\begin{equation}\label{eq:channel_matrix_extended}
		\mathbf{H}_e(\alpha_e) = [\sqrt{P_1}\mathbf{a}_1^{\text{ESG}}(\theta_1, r_1, \alpha_e), \dots, \sqrt{P_K}\mathbf{a}_K^{\text{ESG}}(\theta_K, r_K, \alpha_e)].
	\end{equation}
	The received signal can thus be expressed as
	\begin{equation}\label{eq:channel_matrix}
		\mathbf{x}_e(t) = \mathbf{H}_e(\alpha_e)\mathbf{s}(t) + \mathbf{n}_e(t).
	\end{equation}
	For subsequent processing, this is written in the equivalent compact form:
	\begin{equation}\label{eq:extended_signal_compact}
		\mathbf{x}_e(t) = \mathbf{A}_{\text{ESG}}(\theta, r, \alpha_e) \tilde{\mathbf{s}}(t) + \mathbf{n}_e(t),
	\end{equation}
	where $\tilde{\mathbf{s}}(t)$ is the effective source signal vector defined in \eqref{eq:effective_signal}. The array manifold matrix $\mathbf{A}_{\text{ESG}}(\theta, r, \alpha_e) \in \mathbb{C}^{M \times K}$ is constructed from the exact ESG steering vectors:
	\begin{equation}\label{eq:esg_manifold_extended}
		\mathbf{A}_{\text{ESG}}(\theta, r, \alpha_e) = [\mathbf{a}_1^{\text{ESG}}(\theta_1, r_1, \alpha_e), \dots, \mathbf{a}_K^{\text{ESG}}(\theta_K, r_K, \alpha_e)].
	\end{equation}
	This high-fidelity model is essential for the joint DoA and range refinement performed in the second stage of the proposed algorithm.

	\begin{remark}
		The key insights from this comparison are threefold: (1) \textbf{Geometric Evolution}: The S-FAS model represents a natural evolution from far-field (geometry-independent) to near-field (geometry-dependent) to scalable (geometry-adaptive) steering vectors, where each level adds complexity but significantly enhances modeling capability. (2) \textbf{Diversity Generation}: Unlike conventional approaches that are constrained by fixed array geometry, the S-FAS model generates multiple geometric perspectives of identical sources through the scaling parameter $\alpha$, effectively transforming a single-configuration estimation problem into a multi-configuration optimization that exploits spatial diversity. (3) \textbf{Unified Framework}: The S-FAS formulation provides a unified mathematical framework that seamlessly handles mixed-field scenarios without requiring prior source classification, as the exact spatial geometry remains valid across all ranges while the scaling parameter adapts the resolution characteristics to match estimation requirements.
	\end{remark}

				\vspace{-3mm}
	\subsection{Data Collection and Covariance Estimation}
	
	To enable statistical signal processing, the S-FAS system collects $N$ temporal snapshots in each configuration.
	
	For the compressed configuration ($\alpha_c$), the data matrix is constructed as:
	\begin{equation}\label{eq:data_matrix_c}
		\mathbf{X}_c = [\mathbf{x}_c(1), \mathbf{x}_c(2), \ldots, \mathbf{x}_c(N)] = \mathbf{H}_c(\alpha_c)\mathbf{S} + \mathbf{N}_c,
	\end{equation}
	where $\mathbf{S} = [\mathbf{s}(1), \mathbf{s}(2), \ldots, \mathbf{s}(N)] \in \mathbb{C}^{K \times N}$ is the signal matrix, and $\mathbf{N}_c \in \mathbb{C}^{M \times N}$ is the noise matrix.
	
	For the extended configuration ($\alpha_e$), the data matrix becomes:
	\begin{equation}\label{eq:data_matrix_e}
		\mathbf{X}_e = [\mathbf{x}_e(1), \mathbf{x}_e(2), \ldots, \mathbf{x}_e(N)] = \mathbf{H}_e(\alpha_e)\mathbf{S} + \mathbf{N}_e.
	\end{equation}
	
	The sample covariance matrices, which are essential for eigenvalue-based spectral estimation, are then computed from these data matrices.
	
	For the compressed configuration, the sample covariance matrix is given by
	\begin{equation}\label{eq:covariance_c}
		\mathbf{R}_c = \frac{1}{N}\mathbf{X}_c\mathbf{X}_c^H = \mathbf{H}_c(\alpha_c)\mathbf{R}_s\mathbf{H}_c^H(\alpha_c) + \sigma^2_n\mathbf{I}_M,
	\end{equation}
	where $\mathbf{R}_s$ is the signal covariance matrix.
	
	For the extended configuration, it is expressed as
	\begin{equation}\label{eq:covariance_e}
		\mathbf{R}_e = \frac{1}{N}\mathbf{X}_e\mathbf{X}_e^H = \mathbf{H}_e(\alpha_e)\mathbf{R}_s\mathbf{H}_e^H(\alpha_e) + \sigma^2_n\mathbf{I}_M.
	\end{equation}
	
	These two covariance matrices, $\mathbf{R}_c$ and $\mathbf{R}_e$, encapsulate the statistical information from the two distinct spatial configurations and form the basis for the subsequent two-stage estimation algorithm.
	
				\vspace{-3mm}
	
	\section{Proposed Two-Stage DOA Estimation with Mutual Coupling Mitigation}\label{sec:algorithms}
	To highlight the advantages of the S-FAS architecture, this section re-formulates the  DoA estimator based on the two-stage signal model developed in Sections~III V.  In contrast to the legacy three-stage framework preserved in the previous section, the new estimator consists of only \textbf{Stage~1 (compressed configuration)} and \textbf{Stage~2 (extended configuration)}.  For ease of comparison, original notation is retained whenever possible.
	
	\subsection{Stage 1: Initial DoA Estimation from Compressed Configuration}
	To mitigate the strong edge effects of mutual coupling while maintaining computational efficiency, Stage 1 employs a spatial smoothing technique by selecting a central subarray. This is achieved by applying a selection matrix $\mathbf{F} \in \mathbb{R}^{(M-2p) \times M}$ that extracts the central $M-2p$ elements:
	\begin{equation}\label{eq:selection_matrix}
		\mathbf{F} = [\mathbf{0}_{(M-2p) \times p}, \mathbf{I}_{M-2p}, \mathbf{0}_{(M-2p) \times p}].
	\end{equation}
	
	The selection of central elements is not arbitrary but mathematically optimal. When mutual coupling is present, the central $M-2p$ elements exhibit a crucial property: their coupling-affected steering vector elements maintain uniform amplitude profiles and differ only in phase offsets. Specifically, for these central elements:
	\begin{itemize}
		\item The mutual coupling effects are symmetric about the array center, ensuring balanced coupling contributions;
		\item Each element experiences identical coupling from both its left and right neighbors due to the Toeplitz structure;
		\item The amplitude distortion becomes uniform across the selected subarray, preserving the array manifold structure.
	\end{itemize}
	This symmetry ensures that after applying $\mathbf{F}$, the resulting spatial covariance matrix preserves the essential Toeplitz structure as would be obtained from an uncoupled array, differing only in the effective signal covariance matrix. In contrast, selecting boundary elements would introduce asymmetric coupling effects that could severely degrade subspace-based DOA estimation accuracy.
	
	Applying this selection matrix to the signal model from \eqref{eq:compact_signal}, we obtain the reduced-dimension signal vector
	\begin{equation}\label{eq:smoothed_signal}
		\tilde{\mathbf x}_c(t)=\mathbf F\mathbf C_c(\alpha_c) \mathbf A_{\mathrm{FF}}(\theta,\alpha_c)\tilde{\mathbf s}(t)+\mathbf F\mathbf n_c(t).
	\end{equation}
	The corresponding covariance matrix of this spatially smoothed signal, $\tilde{\mathbf{R}}_c$, is obtained by applying the selection matrix to the full covariance matrix $\mathbf{R}_c$ from \eqref{eq:covariance_c}, such that $\tilde{\mathbf{R}}_c = \mathbf{F}\mathbf{R}_c\mathbf{F}^{\mathrm H}$. This operation effectively decouples the mutual coupling from the array manifold, resulting in the well-structured form:
	\begin{equation}\label{eq:smoothed_covariance}
		\tilde{\mathbf R}_c\triangleq\mathbb E\{\tilde{\mathbf x}_c(t)\tilde{\mathbf x}_c^\mathrm H(t)\}=\tilde{\mathbf A}_{\mathrm{FF}}\,\tilde{\mathbf R}_s\,\tilde{\mathbf A}_{\mathrm{FF}}^{\mathrm H}+\sigma_n^2\mathbf I_{M-2p},
	\end{equation}
	where $\tilde{\mathbf A}_{\mathrm{FF}}$ contains the centre $M-2p$ rows of $\mathbf A_{\mathrm{FF}}$. The modified signal covariance matrix $\tilde{\mathbf R}_s$ incorporates all mutual coupling effects through
	\begin{align}\label{eq:gamma_matrix}
		\tilde{\mathbf R}_s =\boldsymbol{\Gamma}\mathbf R_s\boldsymbol{\Gamma}^{\mathrm H},
	\end{align}
	where
	\begin{align}\label{eq:gamma_matrix2}
		\boldsymbol{\Gamma} = \operatorname{diag}\Big( \sum_{p=-P}^{P} c_{|p|} e^{j p \omega_1},\; \ldots,\; \sum_{p=-P}^{P} c_{|p|} e^{j p \omega_K} \Big).
	\end{align}
	A key observation from \eqref{eq:gamma_matrix2} is that \emph{all} mutual coupling coefficients $\{c_{|p|}\}$ are absorbed into the diagonal matrix $\boldsymbol{\Gamma}$, leaving the reduced manifold $\tilde{\mathbf A}_{\mathrm{FF}}$ free of coupling perturbations. This decoupling of mutual coupling effects from the array manifold is crucial for maintaining the accuracy of subspace-based DOA estimation.
	
	\begin{remark}
		The trade-off in this compressed configuration is clear: while the far-field approximation simplifies computation, the strong mutual coupling (due to $\alpha_c < 1$) introduces estimation errors if not properly compensated. However, this strong coupling can also provide robustness against individual element failures, motivating our two-stage approach where initial estimates from this robust-but-coupled configuration are refined in the less-coupled extended mode.
	\end{remark}
	
	To estimate the DoAs, we first perform eigen-decomposition of $\tilde{\mathbf R}_c$ to obtain the signal and noise subspaces:
	\begin{equation}\label{eq:evd_rc}
		\tilde{\mathbf R}_c=\mathbf U_s^{(c)}\boldsymbol\Lambda_s^{(c)}\mathbf U_s^{(c)\mathrm H}+\mathbf U_n^{(c)}\boldsymbol\Lambda_n^{(c)}\mathbf U_n^{(c)\mathrm H},
	\end{equation}
	where $\mathbf U_n^{(c)}$ spans the noise subspace. The MUSIC spectrum is then computed as
	\begin{equation}\label{eq:music_stage1_new}
		P_{\text{MUSIC}}^{(1)}(\theta)=\frac{1}{\tilde{\mathbf a}_{\mathrm{FF}}^{\mathrm H}(\theta,\alpha_c)\,\mathbf U_n^{(c)}\mathbf U_n^{(c)\mathrm H}\,\tilde{\mathbf a}_{\mathrm{FF}}(\theta,\alpha_c)},
	\end{equation}
	with $\tilde{\mathbf a}_{\mathrm{FF}}(\theta,\alpha_c)$ being the centre $M-2p$ rows of $\mathbf a_{\mathrm{FF}}(\theta,\alpha_c)$. The $K$ highest peaks of $P_{\text{MUSIC}}^{(1)}(\theta)$ yield the coarse DoA estimates $\{\hat\theta_k^{(1)}\}_{k=1}^K$, which serve as initial values for the subsequent refinement in Stage 2 \cite{music}.
	
	\subsection{Stage 2: Joint DoA/Range Refinement from Extended Configuration}
	In Stage 2, we leverage the coarse DoA estimates $\{\hat\theta_k^{(1)}\}_{k=1}^K$ obtained from the compressed configuration to perform a high-resolution joint estimation of both DoA and range. This stage operates on the data $\mathbf{x}_e(t)$ collected from the extended configuration ($\alpha_e > 1$), where mutual coupling is negligible.
	
	First, the sample covariance matrix of the extended array is computed from $N$ snapshots:
	\begin{equation}
		\mathbf{R}_e = \frac{1}{N}\sum_{n=1}^N \mathbf{x}_e(t_n)\mathbf{x}_e(t_n)^{\mathrm H}.
	\end{equation}
	Through eigenvalue decomposition of $\mathbf{R}_e$, we obtain the noise subspace matrix $\mathbf{U}_n^{(e)}$, which is given by
	\begin{align}\label{eq:evd_re}
		\mathbf{R}_e = \mathbf{U}_s^{(e)}\boldsymbol\Lambda_s^{(e)}(\mathbf{U}_s^{(e)})^{\mathrm H} + \mathbf{U}_n^{(e)}\boldsymbol\Lambda_n^{(e)}(\mathbf{U}_n^{(e)})^{\mathrm H}.
	\end{align}
	To avoid a computationally prohibitive 2-D search over the entire parameter space, a two-step refinement is performed. First, for each coarse DoA estimate $\hat{\theta}_k^{(1)}$, we perform a 1-D MUSIC spectral search to obtain an initial range estimate. The range-dependent MUSIC spectrum is given by
	\begin{align}\label{eq:range_estimation_music}
		&P_{\text{MUSIC,r}}(r | \hat\theta_k^{(1)}) \nonumber\\
		&= \frac{1}{\mathbf{a}_{\text{ESG}}^{\mathrm H}(\hat\theta_k^{(1)}, r, \alpha_e) \mathbf{U}_n^{(e)} (\mathbf{U}_n^{(e)})^{\mathrm H} \mathbf{a}_{\text{ESG}}(\hat\theta_k^{(1)}, r, \alpha_e)}.
	\end{align}
	The initial range estimate $\hat{r}_k^{\text{init}}$ for the $k$-th source is found by locating the peak of this spectrum, which is formulated as
	\begin{equation}\label{eq:initial_range_estimate}
		\hat{r}_k^{\text{init}} = \arg\max_{r \in [r_{\min}, r_{\max}]} P_{\text{MUSIC,r}}(r | \hat\theta_k^{(1)}).
	\end{equation}
	With the initial pair $(\hat\theta_k^{(1)}, \hat{r}_k^{\text{init}})$ serving as a high-quality starting point, a localized 2D MUSIC search is then conducted to obtain the final refined estimates. The joint optimization problem is formulated as
	\begin{align}\label{eq:joint_optimization_final}
		&\{\hat{\theta}_k^{(2)}, \hat{r}_k^{(2)}\} \nonumber\\
		&= \arg\max_{\theta,r} \frac{1}{\mathbf{a}_{\text{ESG}}^{\mathrm H}(\theta, r, \alpha_e)\mathbf{U}_n^{(e)}(\mathbf{U}_n^{(e)})^{\mathrm H}\mathbf{a}_{\text{ESG}}(\theta, r, \alpha_e)},
	\end{align}
	subject to the following constraints:
	\begin{align}
		|\theta - \hat{\theta}_k^{(1)}| &\leq \Delta\theta, \label{eq:angle_constraint}\\
		|r - \hat{r}_k^{\text{init}}| &\leq \Delta r, \label{eq:range_constraint}
	\end{align}
	where $\Delta\theta$ and $\Delta r$ define the local search windows. This strategy of using initial estimates to guide a localized, high-resolution search ensures both computational efficiency and high estimation accuracy.
				\vspace{-3mm}
	\subsection{Optional: Joint Estimation with Mutual Coupling in Extended Configuration}
	While the baseline Stage 2 procedure assumes negligible mutual coupling for simplicity, this section outlines a more robust, optional refinement for scenarios where residual coupling in the extended configuration ($\mathbf{C}_e$) remains a concern. This method adapts the rank-reduction principle, previously discussed for arbitrary arrays~\cite{JHe22}, to our joint DoA/range estimation problem.
	
	The signal model in the presence of extended coupling is:
	\begin{equation}
		\mathbf{x}_e(t) = \mathbf{C}_e \mathbf{A}_{\text{ESG}}(\theta, r, \alpha_e) \tilde{\mathbf{s}}(t) + \mathbf{n}_e(t).
	\end{equation}
	Following the transformation approach, the coupling-affected steering vector $\mathbf{C}_e \mathbf{a}_{\text{ESG}}(\theta, r, \alpha_e)$ can be rewritten as $\mathbf{T}(\theta, r)\mathbf{c}_e$, where $\mathbf{c}_e$ is the vector of unknown coupling coefficients and $\mathbf{T}(\theta, r)$ is a transformation matrix dependent on both DoA and range.
	
	The orthogonality condition with the noise subspace $\mathbf{U}_n^{(e)}$ implies that for the true parameters $(\theta_k, r_k)$, the matrix $\mathbf{Q}(\theta_k, r_k) = \mathbf{T}(\theta_k, r_k)^{\mathrm H} \mathbf{U}_n^{(e)} (\mathbf{U}_n^{(e)})^{\mathrm H} \mathbf{T}(\theta_k, r_k)$ must be rank-deficient. This allows us to formulate a new 2D MUSIC-like spectrum that is robust to unknown mutual coupling, which can be formulated as
	\begin{align}\label{eq:mc_music_2d}
		&P_{\text{MC-MUSIC}}(\theta, r) \nonumber\\
		&= \frac{1}{\lambda_{\min}\left( \mathbf{T}(\theta, r)^{\mathrm H} \mathbf{U}_n^{(e)} (\mathbf{U}_n^{(e)})^{\mathrm H} \mathbf{T}(\theta, r) \right)},
	\end{align}
	where $\lambda_{\min}(\cdot)$ denotes the minimum eigenvalue.
	
	The final estimation is then performed by a localized search for the peak of this robust spectrum, guided by the initial estimates from Stage 1, which is written as
	\begin{equation}
		\{\hat{\theta}_k^{(2)}, \hat{r}_k^{(2)}\} = \arg\max_{|\theta-\hat\theta_k^{(1)}|\le\Delta\theta, |r-\hat r_k^{\text{init}}|\le\Delta r} P_{\text{MC-MUSIC}}(\theta, r).
	\end{equation}
	This approach provides enhanced accuracy when residual coupling is present, at the cost of a more complex spectral search function in Stage 2.

						\vspace{-3mm}
			\section{Simulation Results}\label{sec:simulations}
			
			In this section, we provide comprehensive numerical results to demonstrate the effectiveness and superiority of the proposed S-FAS based localization framework. Our evaluation methodology employs a  ULA with $M=32$ antenna elements and sets the carrier wavelength to $\lambda=1$ mm across all simulations, ensuring consistency and fair comparison with existing methods. 
			
			The statistical robustness of our analysis is ensured through 1000 Monte-Carlo trials for each data point, providing reliable RMSE estimates. Our experimental design follows a systematic approach: when evaluating RMSE performance against signal-to-noise ratio (SNR) variations, the number of snapshots is fixed at $N=500$ to isolate the effect of noise; conversely, when analyzing snapshot dependency, the SNR is maintained at 0 dB to focus purely on data length requirements.  
			
			For performance benchmarking, we include several state-of-the-art conventional algorithms, all operating with fixed arrays using the standard inter-element spacing of $0.5\lambda$. Throughout our analysis, we employ consistent notation where ACC refers to the angle estimates obtained after our compressed-stage processing (Stage 1), while AAR denotes the refined angle and range estimates from our extended-stage refinement (Stage 2). The theoretical performance bounds are represented by CRB1 and CRB2, corresponding to the Cramér-Rao bounds for arrays with $d=0.5\lambda$ and $d=1\lambda$ spacing, respectively.  
			\vspace{-3mm}
			\subsection{Single-Shot Localization Performance}
\begin{figure*}[t]
				\centering
				\subfigure[Conventional FF-MUSIC ($d=0.5\lambda$)]{
					\includegraphics[width=0.35\linewidth]{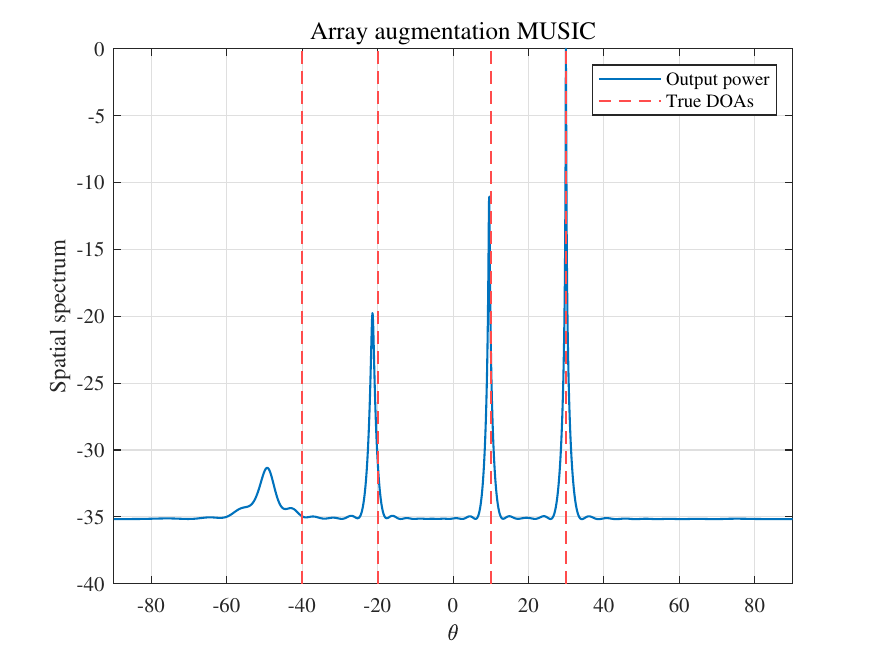}
					\label{fig:doa1}
				}
				\subfigure[Proposed Stage 1 ($d_c=0.1\lambda$)]{
					\includegraphics[width=0.35\linewidth]{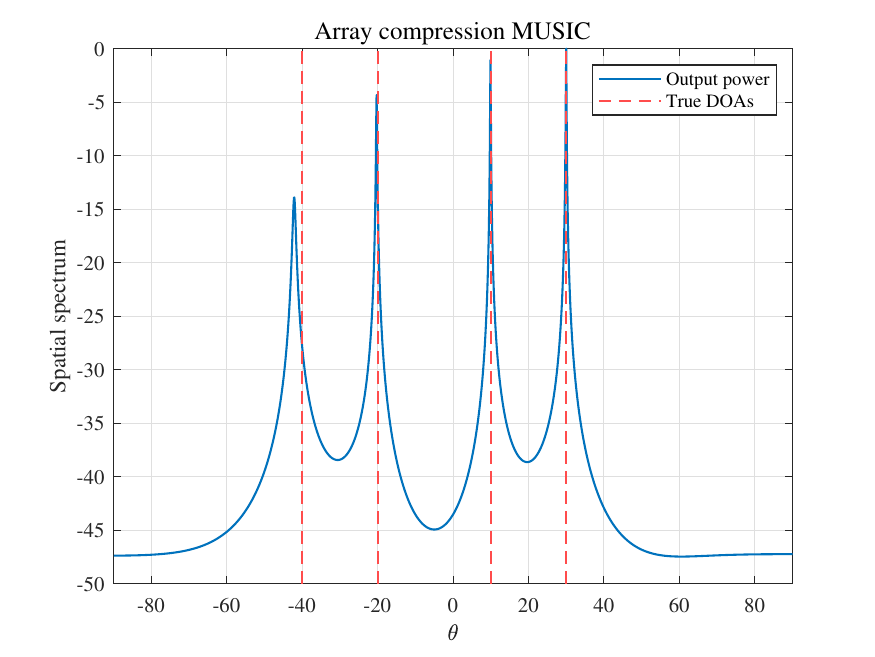}
					\label{fig:doa2}
				}
				\caption{Initial DOA spectral search.}
				\label{fig:doa_estimation}
				\vspace{-5mm}
			\end{figure*}

			We demonstrate the complete single-shot localization process using a challenging mixed-field scenario with four sources at angles $[-40^\circ, -20^\circ, 10^\circ, 30^\circ]$ and distances $[30\lambda, 300\lambda, 1000\lambda, 5000\lambda]$, spanning near-field, Fresnel, and far-field regions. As established in Fig.~\ref{fig:doa_estimation}, our Stage 1 algorithm successfully obtains coarse DOA estimates by compressing the array to $d_c=0.1\lambda$, overcoming the severe model mismatch that causes conventional far-field MUSIC to fail completely.
			
\begin{figure*}[t]
				\centering
				\subfigure[\centering Range search for coarse DOA $\hat{\theta}_1 \approx -40^\circ$]{
					\includegraphics[width=0.22\linewidth]{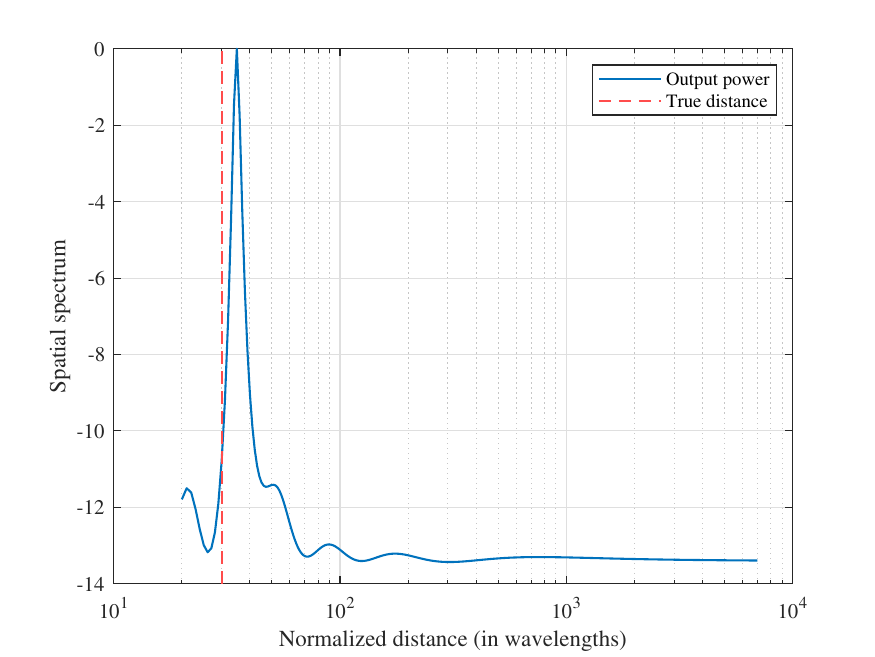}
					\label{fig:range_a}
				}
				\subfigure[\centering Range search for coarse DOA $\hat{\theta}_2 \approx -20^\circ$]{
					\includegraphics[width=0.22\linewidth]{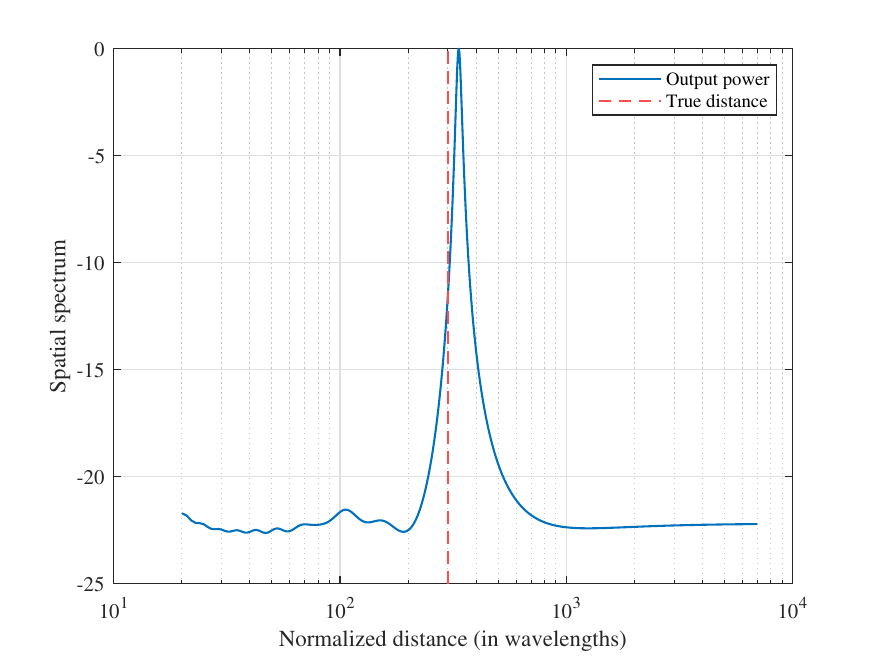}
					\label{fig:range_b}
				}
				\subfigure[\centering Range search for coarse DOA $\hat{\theta}_3 \approx 10^\circ$]{
					\includegraphics[width=0.22\linewidth]{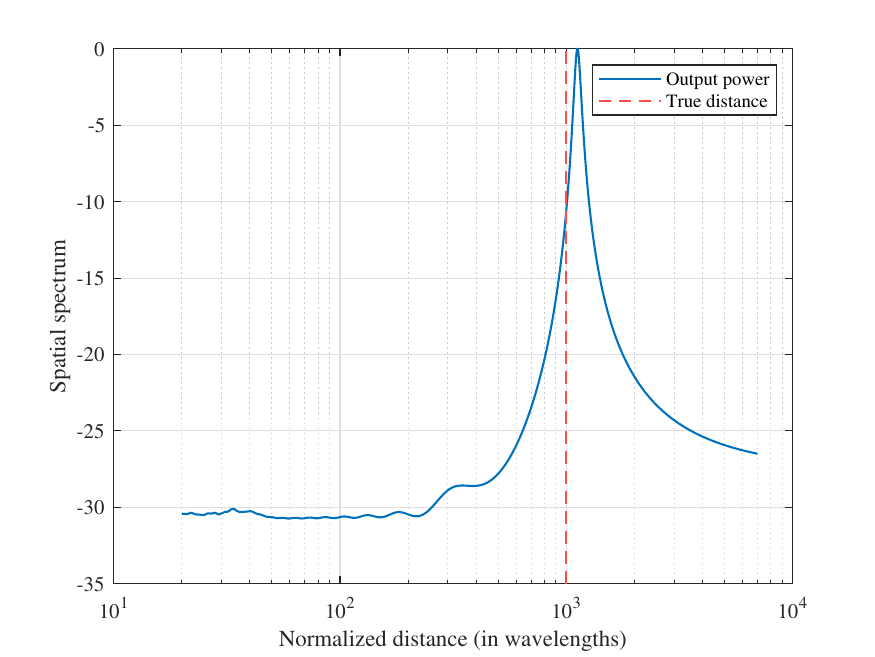}
					\label{fig:range_c}
				}
				\subfigure[\centering Range search for coarse DOA $\hat{\theta}_4 \approx 30^\circ$]{
					\includegraphics[width=0.22\linewidth]{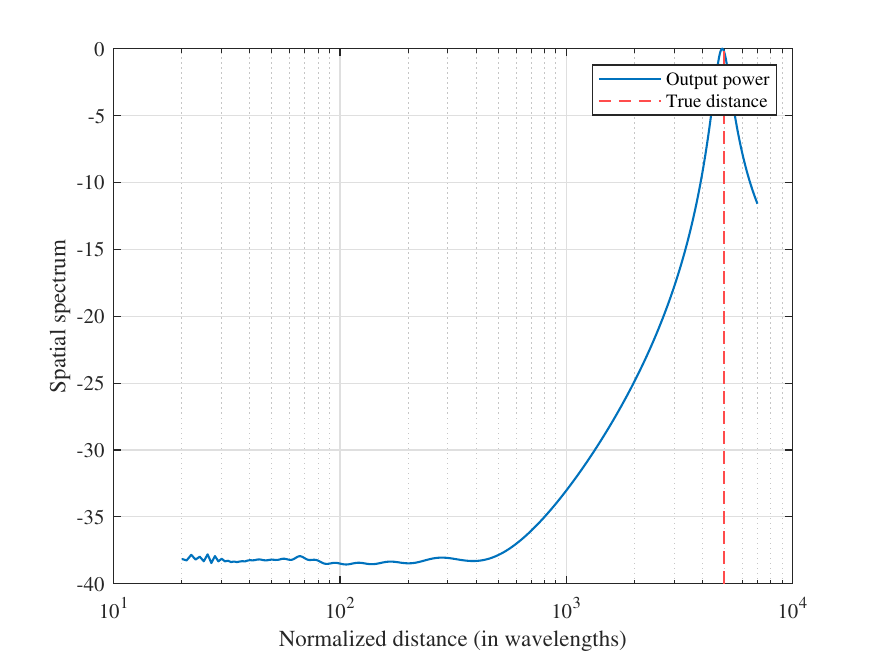}
					\label{fig:range_d}
				}
				\caption{Stage 2: 1D range estimation for each coarse DOA.}
				\label{fig:range_estimation}
					\vspace{-3mm}
			\end{figure*}

			The system transitions to an extended configuration ($d_e=1\lambda$) for precision refinement. The intermediate 1D range search results in Fig.~\ref{fig:range_estimation} demonstrate exceptional performance across all field regions: Fig.~\ref{fig:range_a} shows a sharp peak at 30$\lambda$ for the near-field source, Fig.~\ref{fig:range_b} accurately identifies the Fresnel-region source at 300$\lambda$, while Figs.~\ref{fig:range_c} and \ref{fig:range_d} validate reliable ranging for the far-field sources at 1000$\lambda$ and 5000$\lambda$, respectively. This step successfully provides high-quality initial range estimates $\hat{r}_k^{\text{init}}$ that enable efficient localized 2D refinement.
			
\begin{figure*}[t]
				\centering
				\subfigure[Refined estimate for source 1 ($\theta_1=-40^\circ, r_1=30\lambda$)]{
					\includegraphics[width=0.22\linewidth]{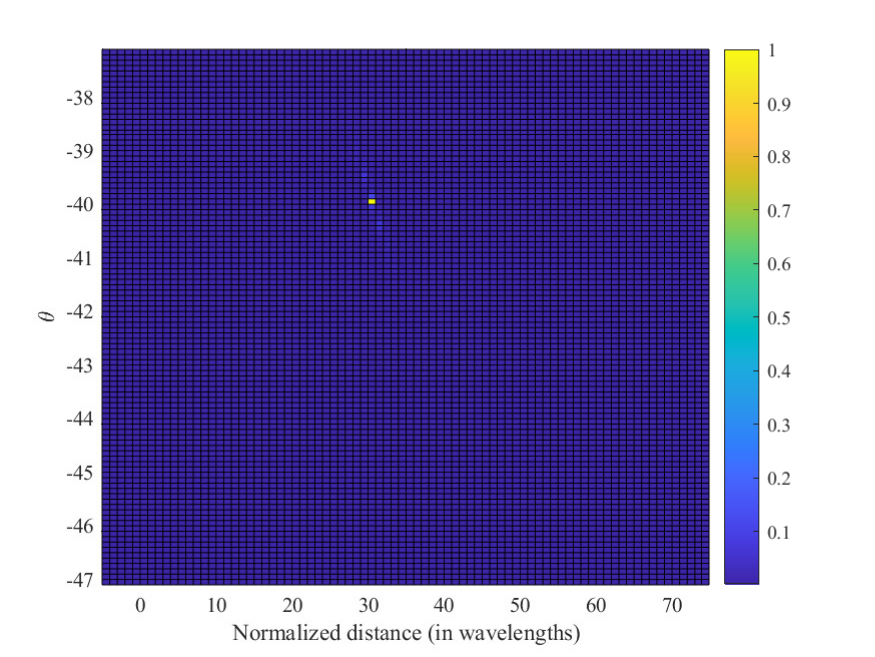}
					\label{fig:2d_a}
				}
				\subfigure[Refined estimate for source 2 ($\theta_2=-20^\circ, r_2=300\lambda$)]{
					\includegraphics[width=0.22\linewidth]{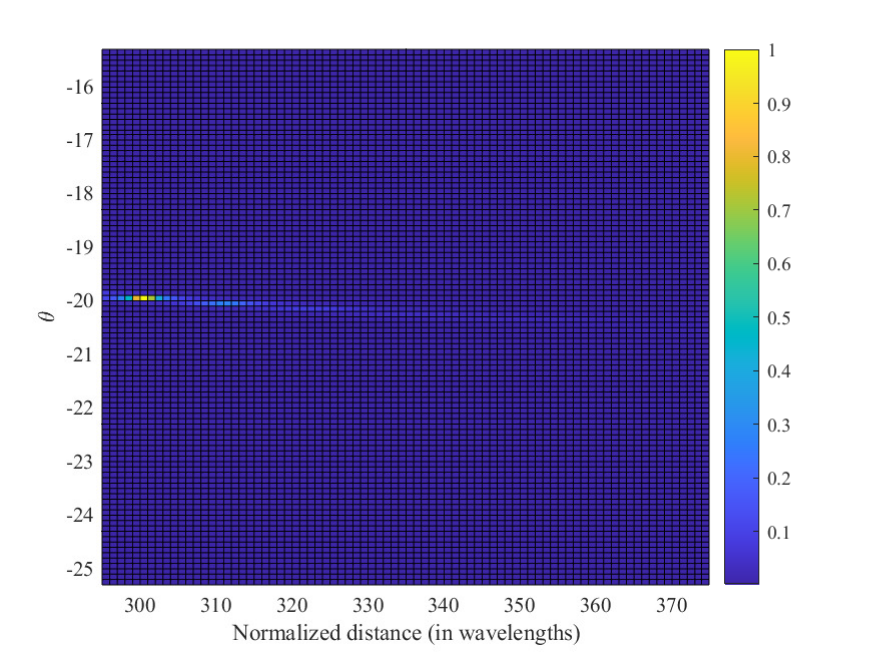}
					\label{fig:2d_b}
				}
				\subfigure[Refined estimate for source 3 ($\theta_3=10^\circ, r_3=1000\lambda$)]{
					\includegraphics[width=0.22\linewidth]{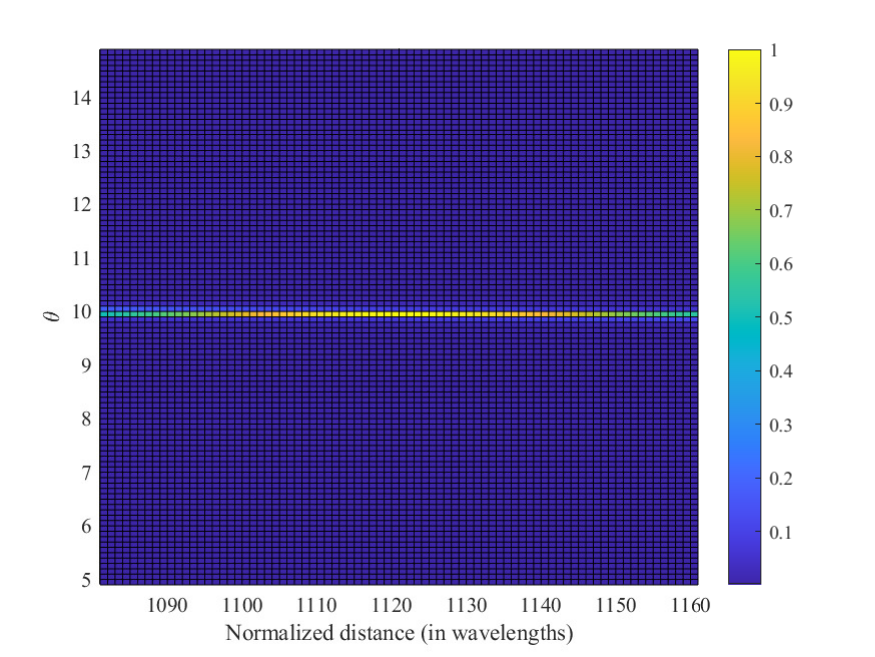}
					\label{fig:2d_c}
				}
				\subfigure[Refined estimate for source 4 ($\theta_4=30^\circ, r_4=5000\lambda$)]{
					\includegraphics[width=0.22\linewidth]{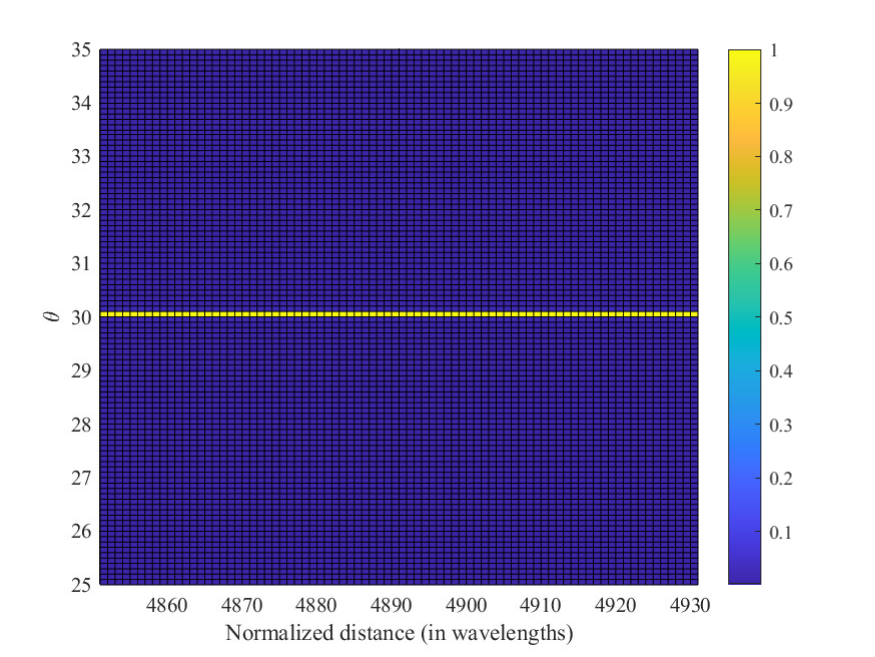}
					\label{fig:2d_d}
				}
				\caption{Stage 2: Final 2D refined DOA and range estimation.}
				\label{fig:2d_estimation}	\vspace{-5mm}
			\end{figure*}

			The final results in Fig.~\ref{fig:2d_estimation} reveal the adaptive behavior of our unified ESG framework. For the near-field source (Fig.~\ref{fig:2d_a}), the algorithm achieves excellent resolution in both dimensions, producing a precise point-like peak. As sources move to greater distances, the spectral characteristics naturally evolve: Fig.~\ref{fig:2d_b} shows elongation along the range axis for the Fresnel source, while Figs.~\ref{fig:2d_c} and \ref{fig:2d_d} demonstrate the graceful transition to high-resolution DOA estimation for far-field sources, where the peaks form sharp horizontal lines. This adaptive behavior validates our method's ability to seamlessly handle the entire spectrum from near-field to far-field scenarios within a single, unified framework, achieving high precision while maintaining computational efficiency through intelligent problem decomposition.
			
			\vspace{-3mm}
			\subsection{RMSE Performance in Far-Field Scenarios}
\begin{figure*}[t]
				\centering
				\subfigure[RMSE vs. SNR]{
					\includegraphics[width=0.45\linewidth]{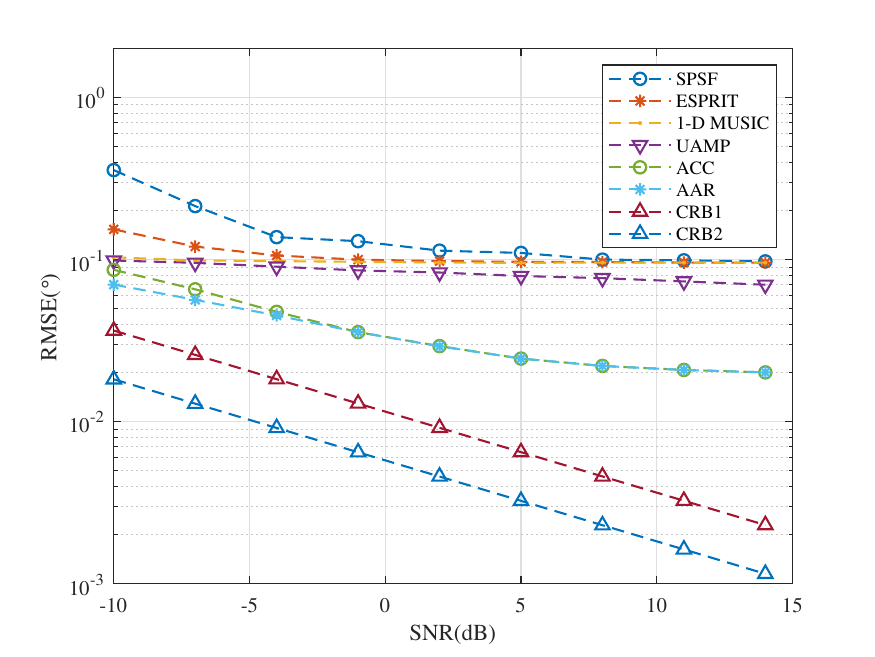}
				}
				\subfigure[RMSE vs. Snapshots]{
					\includegraphics[width=0.45\linewidth]{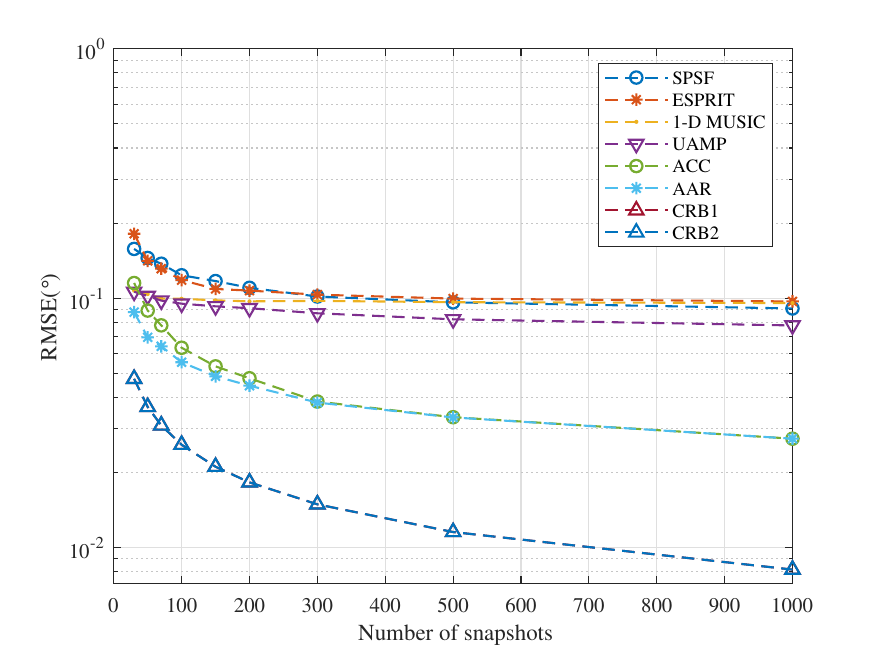}
				}
				\caption{DOA estimation RMSE for two far-field sources.}
				\label{fig:ff_rmse}	\vspace{-5mm}
			\end{figure*}

			We now evaluate the RMSE performance with two far-field sources at $(\theta_1, r_1) = (-20.66^\circ, 4000\lambda)$ and $(\theta_2, r_2) = (10.77^\circ, 5000\lambda)$. Fig.~\ref{fig:ff_rmse} presents a comprehensive comparison of our proposed method against several benchmark algorithms across varying SNR and snapshot conditions.
			
			As demonstrated in Fig.~\ref{fig:ff_rmse}(a), our two-stage approach exhibits remarkable performance characteristics. The initial compressed-array estimates (ACC) provide a robust starting point, maintaining reasonable accuracy even at low SNR values. However, the true strength of our method becomes evident through the refined estimates (AAR), which consistently achieve performance very close to the theoretical CRB2 bound corresponding to the extended array configuration. Notably, both ACC and AAR significantly outperform conventional methods such as ESPRIT and 1-D MUSIC across the entire SNR range, with performance gaps widening substantially at higher SNR values.
			
			The snapshot dependency analysis in Fig.~\ref{fig:ff_rmse}(b) reveals equally compelling results. While traditional methods show gradual improvement with increasing snapshots, our AAR approach demonstrates superior convergence characteristics, rapidly approaching the CRB2 bound with relatively few snapshots. The ACC estimates also show consistent performance, validating the robustness of our compressed-array initialization strategy.
			\vspace{-3mm}
			\begin{remark}
				The results reveal several key insights about our S-FAS framework: (1) \textbf{Graceful Performance Scaling:} The two-stage design ensures that even the initial estimates (ACC) provide reasonable performance, offering a safety net against potential refinement failures. (2) \textbf{Near-Optimal Efficiency:} The AAR performance closely tracks the CRB2 bound, indicating that our algorithm successfully exploits the full potential of the extended array configuration. (3) \textbf{Robustness Advantage:} The substantial performance gap between our method and conventional approaches at high SNR demonstrates that our exact spatial geometry model eliminates the systematic errors inherent in far-field approximations, even for sources that would traditionally be considered ``far-field." This suggests that the benefits of our approach extend beyond mixed-field scenarios to improve performance in conventional applications as well.
			\end{remark}
			
			\vspace{-5mm}
			\subsection{RMSE Performance in Near-Field Scenarios}
\begin{figure*}[t]
				\centering
				\subfigure[RMSE vs. SNR]{
					\includegraphics[width=0.45\linewidth]{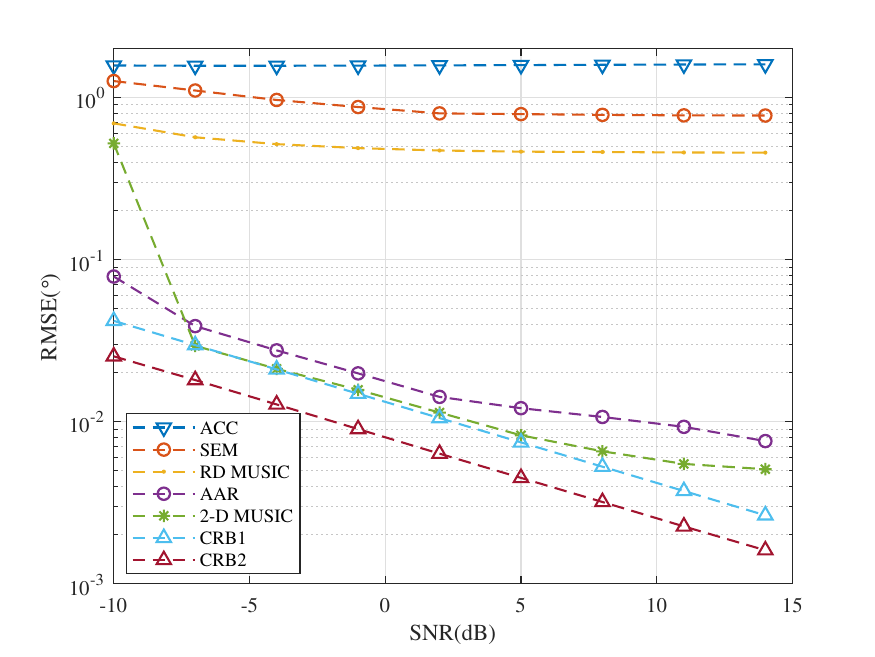}
				}
				\subfigure[RMSE vs. Snapshots]{
					\includegraphics[width=0.45\linewidth]{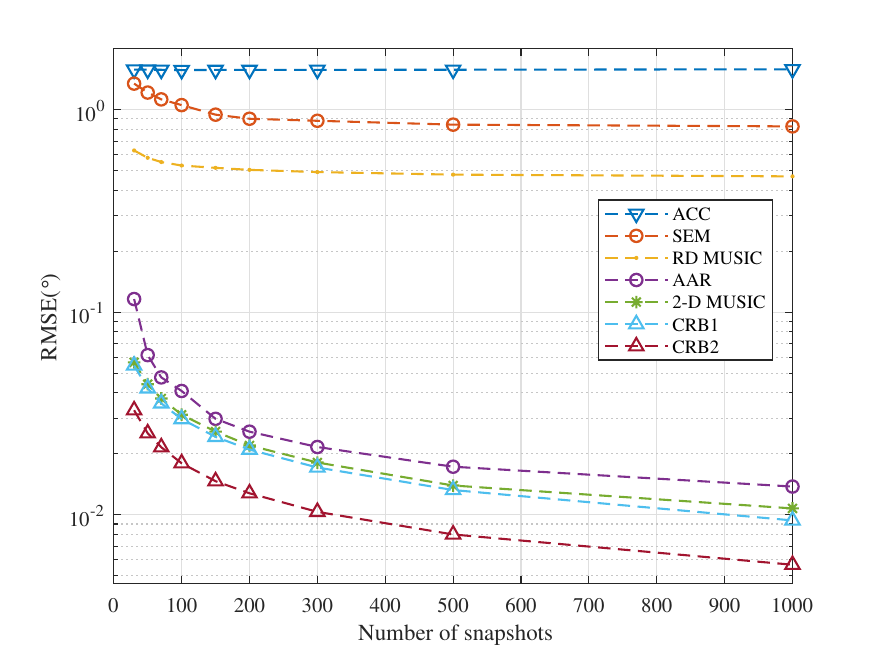}
				}
				\caption{DOA estimation RMSE for two near-field sources.}
				\label{fig:nn_angle_rmse}	\vspace{-5mm}
			\end{figure*}

			Next, we consider a scenario with two near-field sources located at $(\theta_1, r_1) = (-20.66^\circ, 30\lambda)$ and $(\theta_2, r_2) = (10.77^\circ, 200\lambda)$. Fig.~\ref{fig:nn_angle_rmse} presents the DOA estimation RMSE performance, revealing distinctive characteristics that differentiate near-field localization from far-field scenarios.
			
			The SNR analysis in Fig.~\ref{fig:nn_angle_rmse}(a) demonstrates a striking performance hierarchy. Conventional far-field methods (ESPRIT, 1-D MUSIC) exhibit severe performance degradation due to model mismatch, with RMSE values remaining unacceptably high across all SNR levels. In contrast, our compressed-array estimates (ACC) show reasonable performance at moderate to high SNR, while the refined estimates (AAR) achieve exceptional accuracy, closely approaching the CRB2 bound. Notably, the performance gap between our method and conventional approaches is even more pronounced than in far-field scenarios, highlighting the critical importance of exact spatial modeling for near-field sources.
			
			Fig.~\ref{fig:nn_angle_rmse}(b) reveals interesting snapshot dependency characteristics unique to near-field estimation. While conventional methods show limited improvement with increased snapshots due to fundamental model limitations, our AAR approach exhibits rapid convergence to near-optimal performance with relatively few snapshots. The ACC estimates demonstrate consistent behavior, providing reliable initialization across varying data lengths.
			
\begin{figure*}[t]
				\centering
				\subfigure[RMSE vs. SNR]{
					\includegraphics[width=0.45\linewidth]{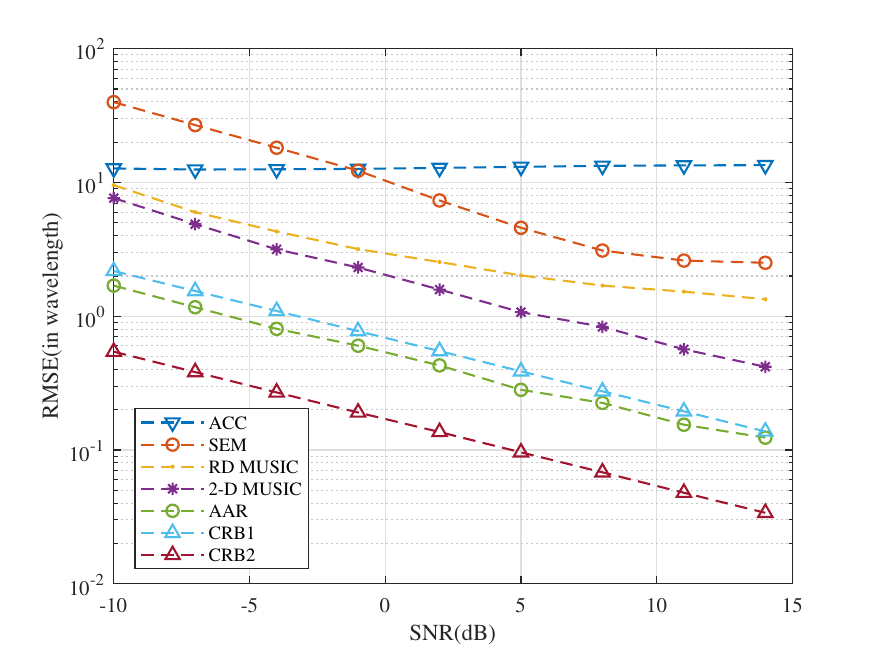}
				}
				\subfigure[RMSE vs. Snapshots]{
					\includegraphics[width=0.45\linewidth]{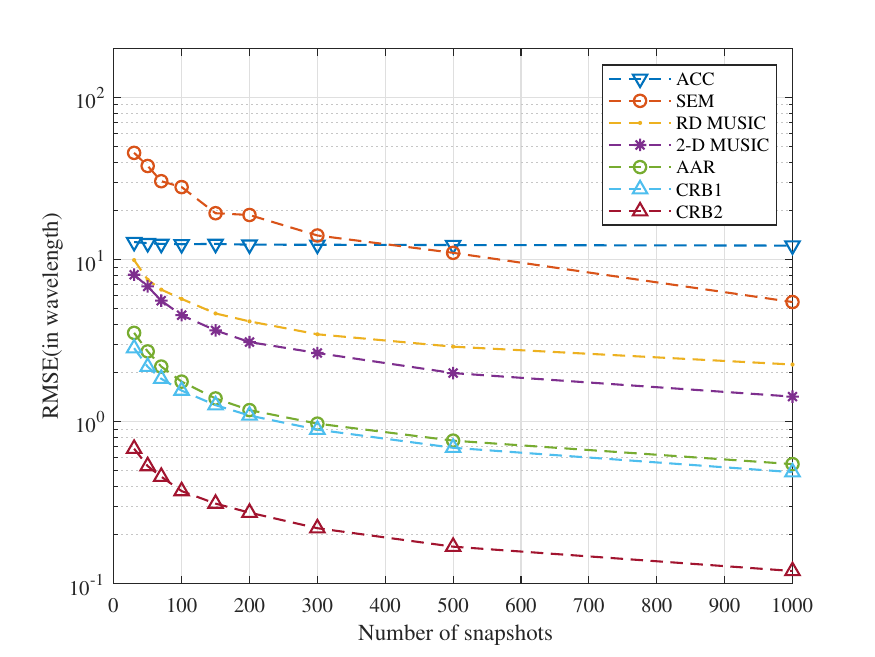}
				}
				\caption{Range estimation RMSE for two near-field sources.}
				\label{fig:nn_range_rmse}	\vspace{-5mm}
			\end{figure*}

			For range estimation performance shown in Fig.~\ref{fig:nn_range_rmse}, our method's superiority becomes even more evident. As anticipated, the AAR performance significantly surpasses the CRB1 baseline (corresponding to conventional half-wavelength arrays), validating the substantial benefits of the extended aperture for range estimation. The proximity to CRB2 confirms that our algorithm effectively exploits the enhanced spatial resolution of the extended configuration.
			
			The detailed analysis of Fig.~\ref{fig:nn_range_rmse} reveals remarkable characteristics specific to range estimation in near-field scenarios. In Fig.~\ref{fig:nn_range_rmse}(a), the SNR dependency shows that conventional methods (RD MUSIC, 2-D MUSIC) struggle significantly with range estimation, exhibiting poor convergence and high error floors. This is expected since these methods rely on far-field approximations that fundamentally cannot capture the range-dependent wavefront curvature essential for accurate distance estimation. In stark contrast, our AAR estimates demonstrate exceptional performance, achieving sub-wavelength range accuracy at moderate SNR levels and closely tracking the CRB2 bound across the entire SNR range.
			
			Fig.~\ref{fig:nn_range_rmse}(b) provides equally compelling evidence of our method's efficiency in terms of data requirements. While conventional approaches show minimal improvement with increased snapshots due to their inherent model limitations, our AAR method rapidly converges to near-optimal performance with remarkably few snapshots. This rapid convergence is particularly valuable in practical applications where data collection time is limited or computational resources are constrained.
			\vspace{-3mm}
			\begin{remark}
				The range estimation results highlight a fundamental advantage of the S-FAS paradigm: \textbf{Unified Range Capability}. Unlike conventional approaches that treat range estimation as a secondary byproduct of angle estimation, our exact spatial geometry model naturally incorporates range information as a first-class parameter. This leads to: (1) \textbf{Dramatic Performance Gains:} The gap between our method and conventional approaches is even larger for range estimation than for angle estimation, demonstrating the critical value of the extended aperture and exact modeling. (2) \textbf{Computational Efficiency:} The two-stage decomposition avoids the computationally burdensome of full 2D searches while achieving near-optimal performance. (3) \textbf{Practical Viability:} The rapid convergence with few snapshots makes our approach suitable for real-time applications where quick and accurate range estimates are essential.
			\end{remark}
			
			\vspace{-5mm}
			\subsection{RMSE Performance in Mixed-Field Scenarios}
			Finally, we test a complex mixed-field scenario with three sources: a near-field source at $(\theta_1, r_1) = (-20.66^\circ, 30\lambda)$, a Fresnel-region source at $(\theta_2, r_2) = (10.77^\circ, 500\lambda)$, and a far-field source at $(\theta_3, r_3) = (30.88^\circ, 5000\lambda)$. The Fresnel distance for this array is $R_F = 512\lambda$. This scenario represents the most challenging test case, as it requires the algorithm to simultaneously handle sources from all three distinct propagation regimes within a single estimation framework.
			
\begin{figure*}[t]
				\centering
				\subfigure[RMSE vs. SNR]{
					\includegraphics[width=0.45\linewidth]{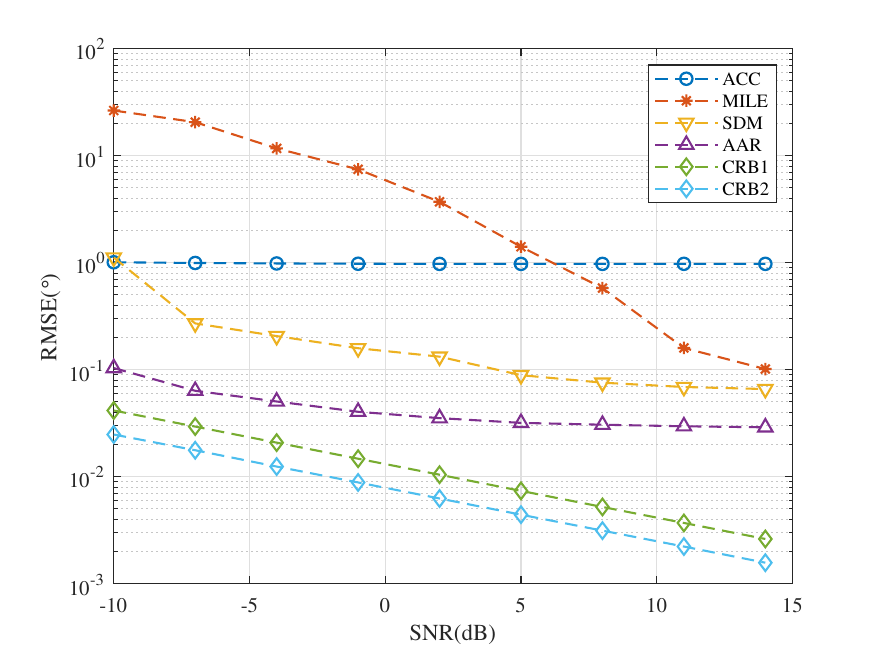}
				}
				\subfigure[RMSE vs. Snapshots]{
					\includegraphics[width=0.45\linewidth]{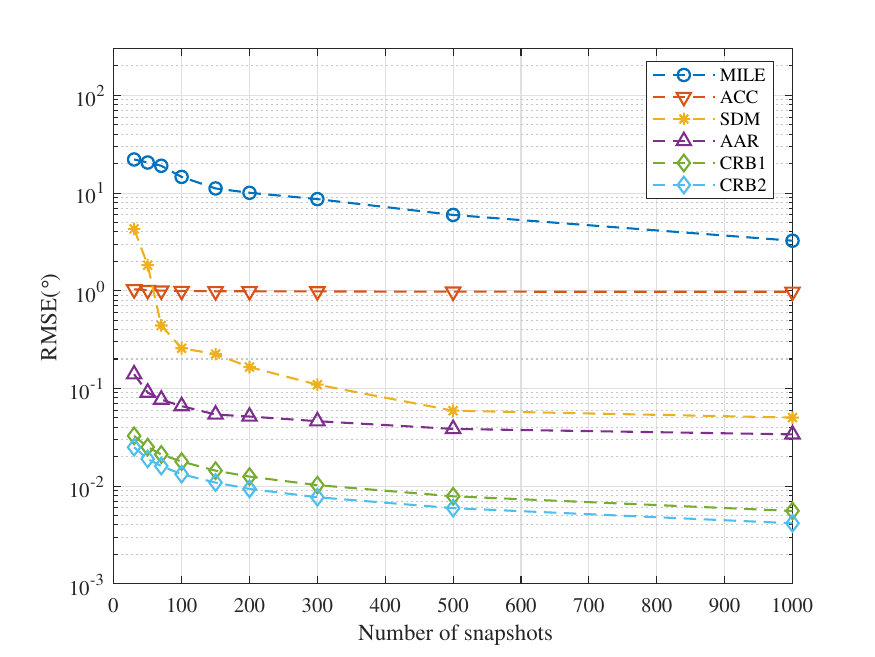}
				}
				\caption{DOA estimation RMSE for mixed-field sources.}
				\label{fig:nf_angle_rmse}	\vspace{-5mm}
			\end{figure*}

			The DOA estimation results in Fig.~\ref{fig:nf_angle_rmse} demonstrate the remarkable robustness of our unified approach. Fig.~\ref{fig:nf_angle_rmse}(a) reveals that conventional methods exhibit catastrophic failure in this mixed-field environment, with some algorithms (MULI) showing complete breakdown across all SNR levels. Other traditional methods (SDM) show modest performance but remain far from optimal. In striking contrast, our method maintains exceptional performance consistency: the ACC estimates provide reliable initialization, while the AAR refinement achieves performance very close to the CRB2 bound for all three sources simultaneously, regardless of their diverse field classifications.
			
			The snapshot analysis in Fig.~\ref{fig:nf_angle_rmse}(b) further validates our method's practical viability. While conventional approaches show erratic behavior and poor convergence characteristics in this complex scenario, our AAR method demonstrates smooth and rapid convergence to near-optimal performance across all source types with minimal data requirements.
			
\begin{figure*}[t]
				\centering
				\subfigure[RMSE vs. SNR]{
					\includegraphics[width=0.45\linewidth]{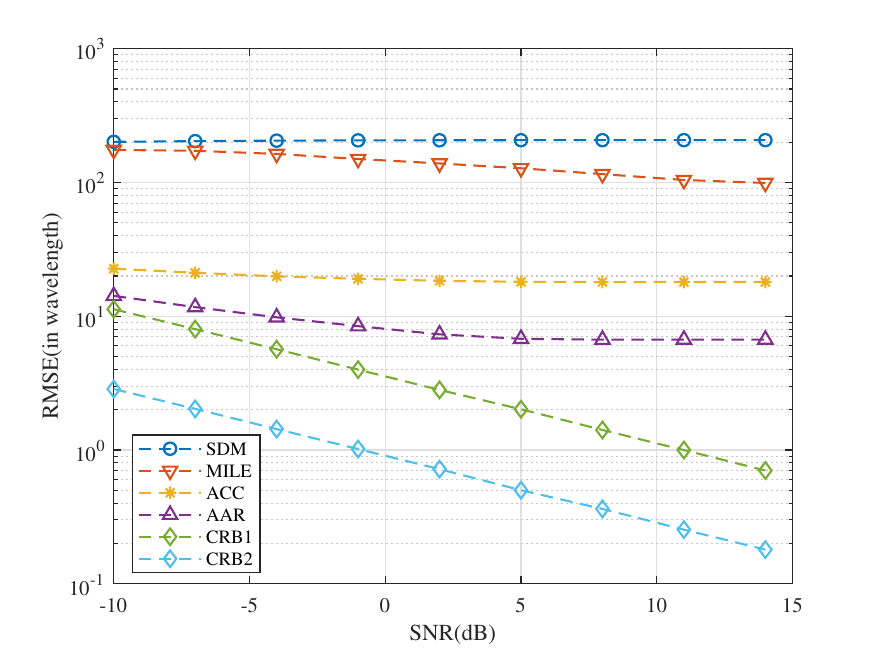}
				}
				\subfigure[RMSE vs. Snapshots]{
					\includegraphics[width=0.45\linewidth]{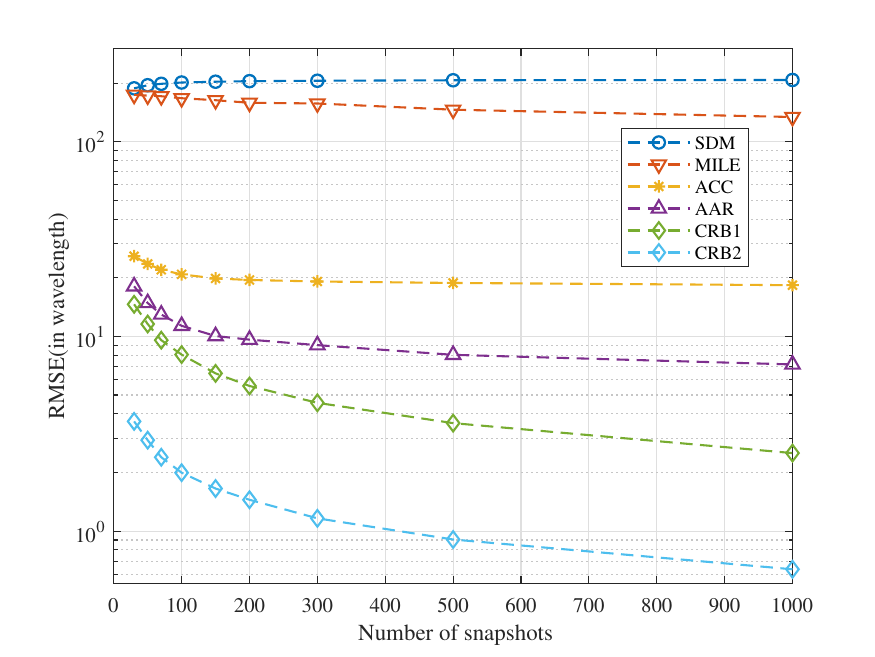}
				}
				\caption{Range estimation RMSE for mixed-field sources.}
				\label{fig:nf_range_rmse}	\vspace{-5mm}
			\end{figure*}

			For range estimation in this mixed-field scenario, Fig.~\ref{fig:nf_range_rmse} presents the most compelling evidence of our framework's superiority. The performance analysis focuses on the near-field and Fresnel-region sources (since the far-field source provides minimal range information). Fig.~\ref{fig:nf_range_rmse}(a) shows that conventional methods completely fail to provide meaningful range estimates, exhibiting extremely high error levels that remain constant across SNR variations. This failure stems from their inability to properly model the mixed propagation characteristics. Our AAR method, however, achieves remarkable range estimation accuracy for both the near-field (30$\lambda$) and Fresnel-region (500$\lambda$) sources, tracking the CRB2 bound closely and demonstrating sub-wavelength precision at high SNR. The convergence behavior in Fig.~\ref{fig:nf_range_rmse}(b) reinforces the practical advantages of our approach. Even in this challenging mixed-field environment, our method requires relatively few snapshots to achieve near-optimal performance, making it suitable for real-time applications where rapid and accurate localization across diverse source types is essential.
			\vspace{-3mm}
			\begin{remark}
				The mixed-field results provide the ultimate validation of our S-FAS paradigm's core philosophy: \textbf{Universal Adaptability}. The key insights are: (1) \textbf{Unified Performance:} Our method achieves consistent, near-optimal performance across all field regimes simultaneously, eliminating the need for pre-classification or field-specific algorithms. (2) \textbf{Graceful Scalability:} Unlike conventional methods that fail catastrophically in mixed scenarios, our approach maintains robust performance regardless of source diversity. (3) \textbf{Real-World Viability:} The combination of excellent accuracy, rapid convergence, and computational efficiency makes our framework uniquely suited for practical applications where source field characteristics are unknown or variable.  
			\end{remark}
			
			\vspace{-5mm}
			
			\section{Conclusion}\label{sec:conclusion}
			In this paper, we presented the S-FAS framework for high-precision source localization in mixed-field scenarios. The key innovation lies in a two-stage estimation strategy that leverages array reconfigurability: a compact configuration for robust initial DOA estimation, followed by an extended configuration for precise range refinement. Underpinned by an exact spatial geometry model, this approach removes the traditional need for source field classification while maintaining near-optimal performance across diverse propagation regimes. Comprehensive simulations validated the framework?s effectiveness, achieving angular accuracy near theoretical bounds and sub-wavelength range estimation precision. Future research will address current limitations by: (1) developing robust tracking algorithms to handle source mobility during configuration transitions, and (2) extending from 2D to full 3D positioning through elevation angle estimation with planar or volumetric S-FAS arrays.
			

		\end{document}